\documentclass[aps,pra,showpacs,footinbib,twoside,twocolumn,10pt]{revtex4-2}
\usepackage[colorlinks=true, citecolor=blue, urlcolor=blue ]{hyperref}
\usepackage{epsfig,newlfont,amssymb,amsfonts,amsmath,bm,subfigure,palatino,mathtools,amsthm,braket,soul,enumitem,graphics,graphicx,times,physics, multirow, makecell}
\usepackage[normalem]{ulem}
\usepackage{tabularx}
\usepackage[table,xcdraw]{xcolor}

\newcommand{\adi}[1]{{\color{magenta}{{#1}}}}

\begin{document}

\title{
Spread complexity as a probe in generalized and long-range Aubry--Andr\'e--Harper models
}

\author{Triyas Sapui$^{1,2}$, Tanoy Kanti Konar$^{1,2,3}$, Subinay Dasgupta$^{1}$ and Aditi Sen(De)$^{1,2}$  }

\affiliation{$^1$Harish-Chandra Research Institute,  Chhatnag Road, Jhunsi, Prayagraj - 211019, India}
\affiliation{$^2$Homi Bhabha National Institute,  Training School Complex, Anushakti Nagar, Mumbai 400 094, India}
\affiliation{$^3$Instytut Fizyki Teoretycznej, Wydzia\l{} Fizyki, Astronomii i Informatyki Stosowanej, Uniwersytet Jagiello\'nski, \L{}ojasiewicza 11, PL-30-348 Krak\'ow, Poland}
\begin{abstract}

We investigate the spread complexity of quantum quenches in generalized and long-range Aubry--Andr\'e--Harper (AAH) models, encompassing regimes with and without mobility edges. In particular, 
in the generalized AAH  models supporting energy-dependent mobility edges, we demonstrate that the long-time averaged spread complexity exhibits nonanalytic behavior when the post-quench quasiperiodic potential crosses the mobility edge associated with the energy of the initial eigenstate, thereby accurately identifying the mobility-edge transition. Such a behavior is supported by the spreading of local density of states. We further derive analytical expressions for the moments and the corresponding Lanczos coefficients for quenches between the limits of vanishing and strong  quasiperiodic potentials. The Lanczos coefficients display qualitatively distinct behavior depending on the presence of mobility edges --
they exhibit an initial plateau followed by a decay with the Krylov basis index, in contrast to the nearly constant behavior of the conventional AAH model without mobility edges. For LR hopping, the coefficients decay with the Krylov basis index for quenches from the localized to the extended phase, while they coincide with the AAH results for quenches in the opposite direction.

\end{abstract}

\maketitle


\section{Introduction}

Recent advances in the theoretical and experimental development of quantum simulators have opened new avenues for exploring the dynamics of complex quantum many-body systems. A central concept in this context is quantum complexity, which quantifies the difficulty of simulating a quantum state on a quantum device. More precisely, it can be understood as the minimum number of elementary operations or simple building blocks required to prepare a target quantum state from a chosen reference state, making it inherently  context-dependent. Over the past decade, several measures of quantum complexity have been proposed, including Nielsen complexity~\cite{Nielsen2006,Dowling2008}, Kolmogorov complexity~\cite{Kolmogorov1998}, quantum nonstabilizerness~\cite{leone_prl_2022}, each arising in different contexts of quantum information and computation.

Among the various measures, Krylov complexity has recently emerged as a powerful framework for characterizing the growth of quantum operators and states~\cite{Nandy2025}. The operator growth hypothesis states that, in chaotic systems, the Lanczos coefficients should grow as rapidly as possible, with the maximal behavior being linear in the Krylov basis~\cite{Parker2019}. This work has sparked extensive studies of operator growth and Krylov complexity in a variety of settings~\cite{Avdoshkin2020,Barbn2019,Rabinovici2021,Rabinovici2022_a,Rabinovici2022_b} and  diverse physical systems, including the the Sachdev-Ye-Kitaev model~\cite{Jian2021,Menzler_prb_2024}, conformal field theories~\cite{Dymarsky2021}, Floquet circuits~\cite{nizami_pre_2024,Suchsland2025,Streamlined_2025}, open quantum systems~\cite{Bhattacharya2022,Bhattacharya2023}, and systems exhibiting ergodicity breaking and constrained systems~\cite{Menzler2024,Malik_prb_2026}. The concept is subsequently  extended from operators to quantum states~\cite{Balasubramanian2022}, where the Krylov basis is shown to minimize the spread of the wave function. The corresponding measure of state growth, termed as \emph{spread complexity}, has since been  employed to investigate the dynamics of quantum states in a wide range of systems, from single qubit~\cite{Seetharaman_prd_2025} to many-body systems~\cite{Bento2024,Baggioli2025,Balasubramanian2025,Bhattacharjee2025,Bhattacharya2024,Caputa2022,Bhattacharjee_prb_2022,Takahashi2025,Caputa2023,Reza2025,Balasubramanian2023,mamta_prb_2024,Nandy2025,zhou_prr_2025}, non-Hermitian systems~\cite{bhattachrya_prb_2024,guerra_prb_2025,Nandy_prb_2025}, and random unitary circuits~\cite{sahu_prb_2026,chaki2026}. More recently, it has been realized that Krylov complexity may depend  nontrivially  on the choice of the initial operator or state~\cite{PG2025}, raising important questions regarding its universality.

Despite these developments, most investigations on spread complexity have  predominantly confined to interacting many-body quantum systems, while the behavior of complexity in single-particle systems remains comparatively little explored~\cite{Peacock2026,HsiuChung2026}. In this work, we address this gap by investigating the spread complexity and the associated Lanczos coefficients in  generalized and long-range (LR) Aubry--Andr\'e--Harper model~\cite{aubrey_andre,Harper1955,Deng2019}. This model has attracted considerable attention over the past decade as it provides a paradigmatic setting for exploring localization phenomena arising from the competition between hopping and a quasiperiodic onsite potential. Also, at the self-dual point, the short-range AAH chain hosts multifractal eigenstates~\cite{Hiramoto_prb_1989}. Depending on the hopping profile and potential strength, it exhibits a rich phase diagram comprising mobility edges, extended,  and localized phases~\cite{Ganeshan2015,Deng2019,Roy2021,Biddle2011,Biddle2009,Biddle2010,Yang2017,Qing2026,Ye2024}. In contrast to interacting quantum spin models, such as the nearest-neighbor and long range $XY$ or Heisenberg models~\cite{Sachdev2011}, the generalized AAH model can be realized by engineering only the hopping amplitudes and onsite potentials, making its experimental implementation comparatively less demanding.  Moreover, it can be experimentally realized using ultracold atoms in optical lattices~\cite{Lye2005,Roati2008,Lahini2009,Schreiber2015}, while its long-range counterpart can be engineered in trapped-ion, Rydberg-atom platforms~\cite{Kim2010,Richerme2014,Labuhn2016} and such system is beneficial for building quantum quantum sensors~\cite{sahoo_pra_2024,sahoo_prb_2025}.

To characterize the dynamical transitions via spread complexity, the initial state is prepared as one of the eigenstates corresponding to a given value of the quasiperiodic potential strength, followed by an instantaneous quench to a final value. We demonstrate that the long-time averaged spread complexity exhibits pronounced kinks whenever the final potential strength crosses the mobility edge corresponding to the energy of the initial state. This signature is in good agreement with the phase boundaries identified from the inverse participation ratio (IPR). Moreover, we find that the nonanalyticity of spread complexity can be explained by examining the spreading of local density of states.  We further establish a direct connection between these nonanalyticities and the Lanczos coefficients. Specifically, for the AAH model, Lanczos coefficients remain constant with increasing Krylov basis index,  while in the generalized AAH model, they decay after an initial plateau of constant values.  
In addition, we derive analytical expressions for the moments from the survival probability for quenches between states located at the extreme points of different phases. 

Going beyond the AAH model, we also examine the LR AAH model with power-law-decaying hopping strength. Interestingly, we prove that quenches from the extended to the localized regime (low to high quasiperiodic potential strength) yield moments identical to those of the AAH model. In contrast, for quenches in the opposite direction, namely from the localized to the extended regime, the Lanczos coefficients decay with the Krylov basis index, unlike the AAH case. Moreover, we observe that the spread complexity continues to provide a robust indicator of localization, accurately identifying the mobility edges through the appearance of nonanalyticities in its long-time average, although the IPR no longer approaches unity in the localized phase of the LR AAH model.


This paper is organized as follows. In Sec. \ref{sec:krylov_complexity}, we describe the spread complexity and the Hamiltonian for a single particle in a quasiperiodic potential. In Sec. \ref{sec:mobility_edge_detection}, we calculate the time-averaged spread complexity without mobility edge \ref{sec:aubry_andre} and with mobility edge \ref{sec:gen_aubry_andre}. In \ref{sec:analytical_moments} we calculate analytically the moments for  quench of the quasiperiodic potential strength from very small to very high values and vice versa. In Sec. \ref{sec:long_range_spread_complexity}, we investigate the effects of long-range hoppings.

\section{Spread complexity and Model description}
\label{sec:krylov_complexity}

We briefly here review the concept of spread complexity~\cite{Balasubramanian2022,Bento2024,Baggioli2025,Balasubramanian2025,Bhattacharjee2025,Bhattacharya2024,Caputa2022,Takahashi2025,Caputa2023,Reza2025,Balasubramanian2023,Ganguli2024,Scialchi2024,Camargo2024,Grabarits2025,Teh2025}, a useful framework for characterizing the dynamical properties and phases of quantum Hamiltonians, thereby providing a probe to study the dynamical features of quantum many-body systems. It quantifies the extent to which an initial state spreads over the Krylov basis generated through successive applications of the Hamiltonian, commonly referred to as Krylov space~\cite{Balasubramanian2022}. We subsequently introduce a class of long-range Hamiltonians and their phases, and establish Krylov complexity as an efficient dynamical diagnostic of the associated phase transitions.


\subsubsection{Time-averaged spread complexity}

Consider an initial pure state $|\psi(0)\rangle$ evolving under a time-independent Hamiltonian $\hat{H}$, leading to  the time-evolved state,
\begin{equation}
|\psi(t)\rangle=e^{-i\hat{H}t}|\psi(0)\rangle
=\sum_{n=0}^{\infty}\frac{(-it\hat{H})^n}{n!}|\psi(0)\rangle.
\end{equation}
Since the evolved state is generated through repeated action of $\hat{H}$ on the initial state, it can be expressed as a linear combination of the vectors belonging to the Krylov space,
\begin{eqnarray}
    \mathcal{K}= \mathrm{span}\Bigl\{ |\psi(0)\rangle, \hat{H}|\psi(0)\rangle, \hat{H}^{2}|\psi(0)\rangle, \dots\Bigr\}.
    \label{eq:krylovspace}
\end{eqnarray} 
The vectors in $\mathcal{K}$ are generally not orthonormal. Therefore, an orthonormal Krylov basis $\{|K_n\rangle\}$ is constructed using the Lanczos algorithm~\cite{Balasubramanian2022,Nandy2025}, which is equivalent to the Gram-Schmidt orthogonalization procedure~\cite{Balasubramanian2022}. Starting with
\( |K_0\rangle = |\psi(0)\rangle, \qquad |K_{-1}\rangle =0,\) the Lanczos recursion relation is given by
\begin{equation}
|A_{n+1}\rangle
=
\hat{H}|K_n\rangle
-a_n|K_n\rangle
-b_n|K_{n-1}\rangle, n=0, 1, \cdots,
\end{equation}
  where \(a_n=\langle K_n|\hat{H}|K_n\rangle,\) and
\(b_{n+1}=\sqrt{\langle A_{n+1}|A_{n+1}\rangle}.\) The orthonormal basis vectors are then obtained as
\(|K_{n+1}\rangle = \frac{1}{b_{n+1}} |A_{n+1}\rangle\),
provided  $b_{n+1}\neq 0$. The recursion terminates when $b_{n+1}=0$. Because of that, the dimension of the Krylov space is given by \(D_K\) and \(n\leq D_K\). The coefficients $\{a_n,b_n\}$ are known as the Lanczos coefficients and contain valuable information about the dynamical properties of the system, as we will establish in this work. In particular, they have been widely used to distinguish between chaotic and integrable quantum systems~\cite{Baggioli2025,Balasubramanian2025}.

The Lanczos coefficients are closely related to the moments of the survival amplitude, \(S(t)=\langle\psi(0)|\psi(t)\rangle,\) whose derivatives at $t=0$ yield
\begin{equation}
\mu_n = \left. \frac{d^n S(t)}{dt^n}
\right|_{t=0}=
\langle K_0|(i\hat{H})^n|K_0\rangle.
\label{eq:moment_from_survival_probablity}
\end{equation}
These moments can be recursively expressed as functions of the Lanczos coefficients, $\mu_n=f(\{a_m,b_m\})$, and may be efficiently computed using the unwrapped Markov-chain representation which, in turn, can be used to determine phases in dynamics (see Subsec. \ref{sec:analytical_moments}). Having obtained the Lanczos coefficients and the associated Krylov basis, one can define the spread (or Krylov) complexity as
\begin{equation}
\mathcal{C}(t)
=
\sum_{n=0}^{D_K-1}
n\,|\langle K_n|\psi(t)\rangle|^2,
\end{equation}
 Physically, $\mathcal{C}(t)$ measures the average position of the evolving state along the Krylov chain and, therefore, quantifies the extent of state spreading in Krylov space. In this work, our primary quantity of interest is the long-time averaged spread complexity~\cite{Bento2024},
\begin{equation}
\overline{\mathcal{C}}
=
\lim_{t\rightarrow\infty}
\frac{1}{t}
\int_{0}^{t}
\mathcal{C}(t')\dd t'.
\end{equation}
This quantity characterizes the average complexity accumulated during the entire evolution and serves as a useful indicator to probe the dynamical features of the system. Since $\overline{\mathcal{C}}$ depends solely on the Hamiltonian parameters, it provides a convenient tool for identifying different dynamical regimes and equilibrium phase transitions in dynamics, referred to as dynamical quantum phase transitions~\cite{Heyl2018,Bento2024}.

\subsubsection{Long-range and generalized Aubry--Andr\'e--Harper (AAH) Hamiltonian}

To investigate the behavior of spread complexity across the extended-to-localized quantum phase transition, we consider a generalized Aubry--Andr\'e--Harper model with long-range hopping~\cite{Deng2019,Roy2021}. The Hamiltonian reads as 
\begin{eqnarray}
    H = -t \sum_{j< k}^{N} \left(  \frac{c_j^\dagger c_{k}}{|j-k|^\alpha} + \mathrm{h.c.} \right) + \frac{\lambda'  \cos(2\pi q j + \phi)c_j^\dagger c_{j}}{1 - \beta \cos(2\pi q j + \phi)},
\label{eq:ham}
\end{eqnarray}
where $c_j^\dagger$ ($c_j$) denotes the fermionic creation (annihilation) operator at site $j$, $q=(\sqrt{5}-1)/2$ is an irrational number (golden ratio) determining the quasiperiodic modulation, $ \lambda'$ represents the strength of the quasiperiodic potential, we set \(\lambda = \lambda'/t\) to make it dimensionless,  $\alpha$ controls the range of hopping through an algebraic decay of the tunneling amplitude and \(N\) is the system-size. The continuous parameter $\beta\in (-1,1)$ deforms the onsite potential \cite{Ganeshan2015} and plays a crucial role in determining the localization properties of the system. In the short-range limit, i.e.,  $\alpha\rightarrow\infty$, a finite value of $\beta$ gives rise to a single-particle mobility edge, resulting in the coexistence of localized and extended eigenstates within the energy spectrum. On the other hand, long-range hopping with finite $\alpha$ can also induce mobility-edge even when \(\beta=0\)  by altering the localization characteristics of the eigenstates~\cite{Biddle2011,Deng2019}. Consequently, the competition between quasiperiodicity and long-range tunnelling leads to a rich phase diagram  -- extended and localized phases as well as mobility edges. Importantly, as mentioned earlier,  realizing  AAH model is possible in several physical platforms like cold atoms in optical lattices~\cite{Shimasaki2022}, and photonic waveguides~\cite{Lahini2009}. 
Our work employs spread complexity as a dynamical probe to characterize the different phases of the long-range AAH model and to identify the presence of mobility edges. 


\section{mobility edge detection using spread complexity}
\label{sec:mobility_edge_detection}

To investigate the mobility edge (ME) of the generalized AAH model through Krylov complexity, we first consider the case $\beta=0$ and $\alpha\rightarrow\infty$, for which the system reduces to the standard Aubry--Andr\'e-Harper (AAH) model, \(\hat{H}(\lambda)\) with NN hopping. In this limit, the model does not host a mobility edge and exhibits a self-dual localization transition at $\lambda=2$. Specifically, all eigenstates are extended for $\lambda<2$, whereas all eigenstates become localized for $\lambda>2$. To study the long-time averaged spread complexity $\overline{\mathcal{C}}$ across different physical regimes, we
consider two distinct initial states: (i) an extended state, chosen as the ground state of the AAH model at $\lambda_{in}=0$, and (ii) a localized state, obtained as the ground state deep in the localized regime (eg. \(\lambda_{in}=100\)).
In case (i), we perform a sudden quench $\lambda_f\neq 0$ (forward quench),  and for case (ii), we choose \(\lambda_f \neq 100\) (backward quench). The resulting time-evolved state can be written as 
\begin{eqnarray}
    |\psi(t)\rangle = \exp\left(-i \hat{H}(\lambda_f) t\right) |\psi(t=0, \lambda_{in})\rangle,
\end{eqnarray}
and we analyze both the time dependence of $\mathcal{C}(t)$ and its long-time average $\overline{\mathcal{C}}$.

\subsection{Krylov complexity in the Aubry--Andr\'e--Harper model} 
\label{sec:aubry_andre}

\begin{figure}  
    \centering \includegraphics[width=\linewidth]{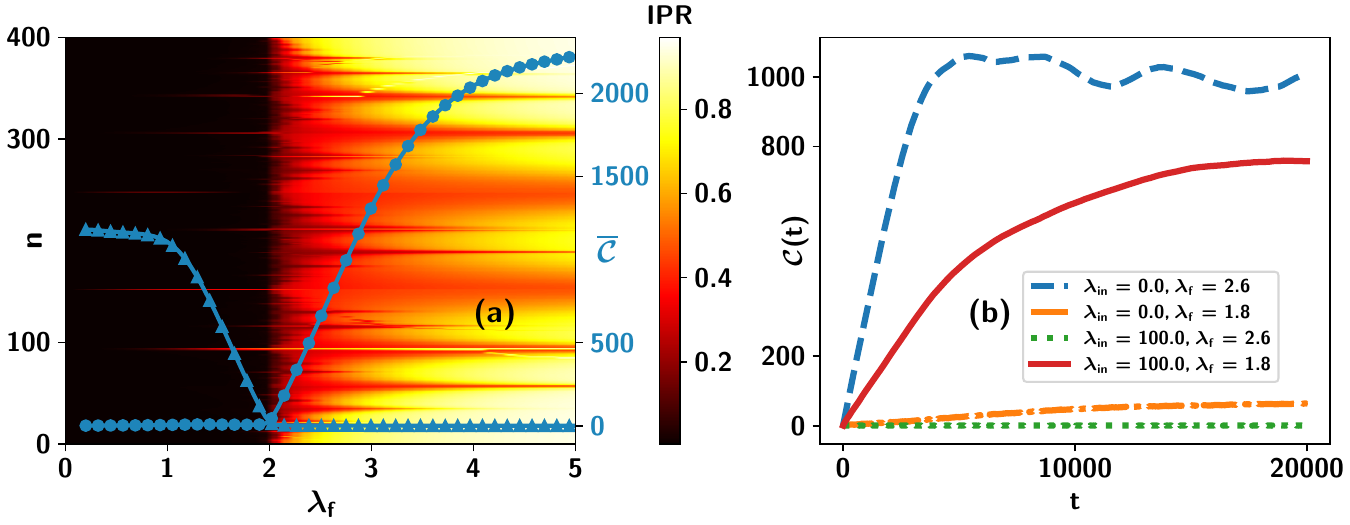}
    \caption{ (a) \textbf{Map plot of the inverse participation ratio (IPR) and the long-time averaged spread complexity (right ordinate) for AAH model.} We plot the spread complexity for the quenches from the ground state of \(H(\lambda_{in}=0)\) to a higher values of \(\lambda_f\) (triangles), and from the ground state of \(H(\lambda_{in}=100)\) to lower values of \(\lambda_f\) (circles) against \(\lambda_f\) (abscissa). The \(x\) axis represents the post-quench quasiperiodic potential strength \(\lambda_f\) and the left \(y\) axis represents the index of the Krylov basis. Other parameters are \(\beta=0\) and \(N=5000\). (b) Variation of \(\mathcal{C}(t)\) (ordinate) with time \(t\) (abscissa) for the same phase quench (dashed-dot  and dotted lines) and a different-phase quench (solid and dashed lnes) for system-size \(N=2000\).  All axes are dimensionless. }
    \label{fig:beta_zero}
\end{figure}

To determine the effectiveness of $\overline{\mathcal{C}}$, we also examine the inverse participation ratio (IPR) of the eigenstates of the AAH model. The IPR of a state $|\psi\rangle$  is defined as~\cite{DomnguezCastro2019} $IPR(|\psi\rangle)=\sum_i |\langle i|\psi\rangle|^4$ for a given state \(|\psi\rangle\) and \(|i\rangle\) represents the state, where the particle is localized at site \(i\). We compute IPR by considering all the eigenstates of the Hamiltonian, \(\hat{H}(\lambda)\), i.e., \(|\psi\rangle \equiv |\psi^r\rangle\) with \(|\psi^r\rangle\) being the eigenvector of  \(\hat{H}(\lambda)\) for a fixed \(\lambda\) (\(r=0,1, \ldots, N-1\)). At first, we calculate the \(C(t)\) for two kinds of initial states, (i) extended and (ii) localized state. We find that when pre- and post-quench parameters belong to the same phase, the spread complexity increases slightly with time, while its huge increment with time is observed when quenching is performed across the phases (see Fig. \ref{fig:beta_zero}(a)).  



{\it Detection of criticality through forward and backward quenches.} 
 When \(\lambda<2\), IPR vanishes, while a finite IPR is obtained when \(\lambda>2\) (see map plot of Fig.~\ref{fig:beta_zero}(a)).
 The simulations with Krylov complexity reveal that for the initial state chosen from the extended phase, $\overline{\mathcal{C}}$ remains nearly zero throughout the extended phase ($\lambda<2$), exhibits a sharp increase at the critical point $\lambda=2$, and subsequently saturates to a finite value in the localized phase, thereby determining the phase boundary (see Fig. \ref{fig:beta_zero}(b)). In contrast, for the backward quench, i.e., for the localized initial state (say, \(\lambda_{in} =100\)),  $\overline{\mathcal{C}}$ assumes a finite value in the extended phase, decreases abruptly at $\lambda=2$, and remains close to zero for $\lambda>2$. Hence, the observed behavior of $\overline{\mathcal{C}}$ is consistent with that of the IPR, thereby confirming Krylov complexity as a reliable dynamical probe of the localization--delocalization transition. This behavior arises because the spread complexity grows when quenching is performed across the phase, while its growth is suppressed for the quench in the same phase, as shown in Fig.~\ref{fig:beta_zero}(b). 
\begin{figure}  
    \centering \includegraphics[width=\linewidth]{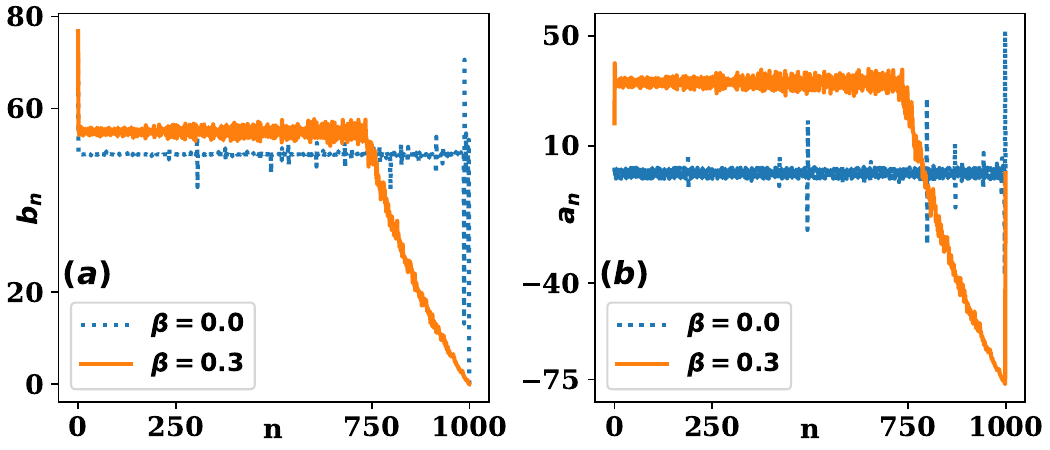}
    \caption{ \textbf{Behavior of the Lanczos coefficients as a function of Krylov basis index \(n\)}. (a) Lanczos coefficients \(b_n\) for the AAH (\(\beta=0.0\)) (dotted) and gAAH (\(\beta=0.3\)) (solid) models. Unlike the AAH model, where \(b_n\) drops to \(0\) abruptly near \(n=D_K =1000\), the gAAH model exhibits a gradual decrease in \(b_n\) after a certain value of \(n\). (b) Lanczos coefficients \(a_n\) for both AAH (\(\beta=0\)) (dotted) an gAAH (\(\beta=0.3\)) (solid) models against \(n\). In the AAH model \(a_n\) fluctuates around \(0\), whereas in the gAAH model, it initially fluctuates around a positive value before gradually decreasing and eventually becoming negative. The data are obtained by quenching the highest excited state of \(\lambda_{in}=0\) to \(\lambda_f=100.0\).  Here \(N=1000\). All axes are dimensionless.}
    \label{fig:bnan_beta0}
\end{figure}

\begin{figure}  
    \centering \includegraphics[width=0.8\linewidth]{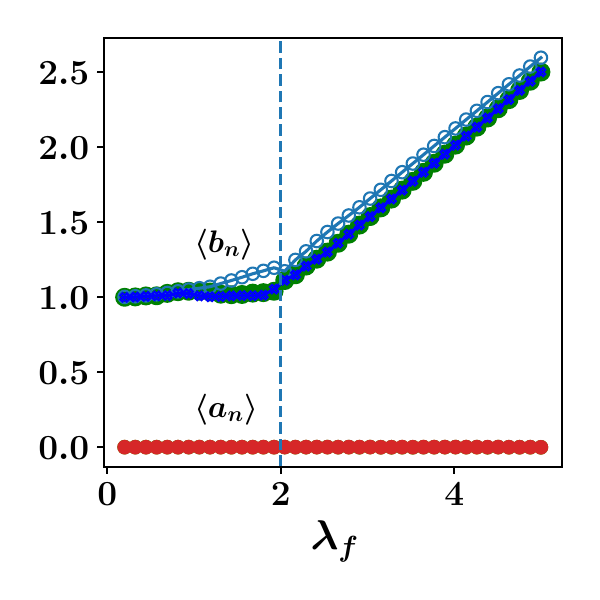}
    \caption{ \textbf{Average of the Lanczos coefficients over \(n\) as a function of \(\lambda_f\) for the AAH model (with \(\beta=0\))}.  The average of \(a_n\) remains close to \(0\) for all values of \(\lambda_f\) for all initial states while  \(\langle b_n \rangle\) is close to \(1\) upto \(\lambda_f=2\) (extended phase) and then it increases linearly in the localized phase for all the initial states, the ground state (solid circles), the eigenstate from the middle spectrum (hollow circles) and the highest excited state (blue stars). \(N =1000\). All axes are dimensionless.}
    \label{fig:avgbnan_beta0}
\end{figure}

To further understand the behavior of Krylov complexity, we analyze the Lanczos coefficients~\cite{Balasubramanian2022}, which provide insight into extended and localized phases.  We observe that although \(b_n\) and \(a_n\) fluctuate with \(n\), they, on average, remain approximately constant  with 
\(n\). 
In order to understand if such Lanczos coefficients carry the information of the phase transition, we compute the average value $\langle b_n\rangle = \frac{1}{D_K} \sum_{n=0}^{D_{K}-1}b_n$ for different initial states and study its dependence on $\lambda_f$. We find that $\langle b_n\rangle$ remains nearly unity throughout the extended phase ($\lambda<2$), but starts growing linearly once $\lambda_f$ exceeds the critical value, $\lambda_f=2$, as shown in Fig. \ref{fig:avgbnan_beta0}. On the other hand, \(\langle a_n\rangle\) remains close to zero for all values of \(\lambda_f\). This observation suggests that the Lanczos coefficients encode valuable information about the underlying phase transition.

The entire analysis of the AAH model confirms that Krylov complexity can be as powerful as IPR. An important question, however, is whether its strength persists in systems with a significantly richer phase diagram than that of the AAH model, which we will answer in the next subsection.

\subsection{Krylov complexity in generalized Aubry-Andr{\'e}-Harper (gAAH)  model}
\label{sec:gen_aubry_andre}

\begin{figure}  
    \centering \includegraphics[width=\linewidth]{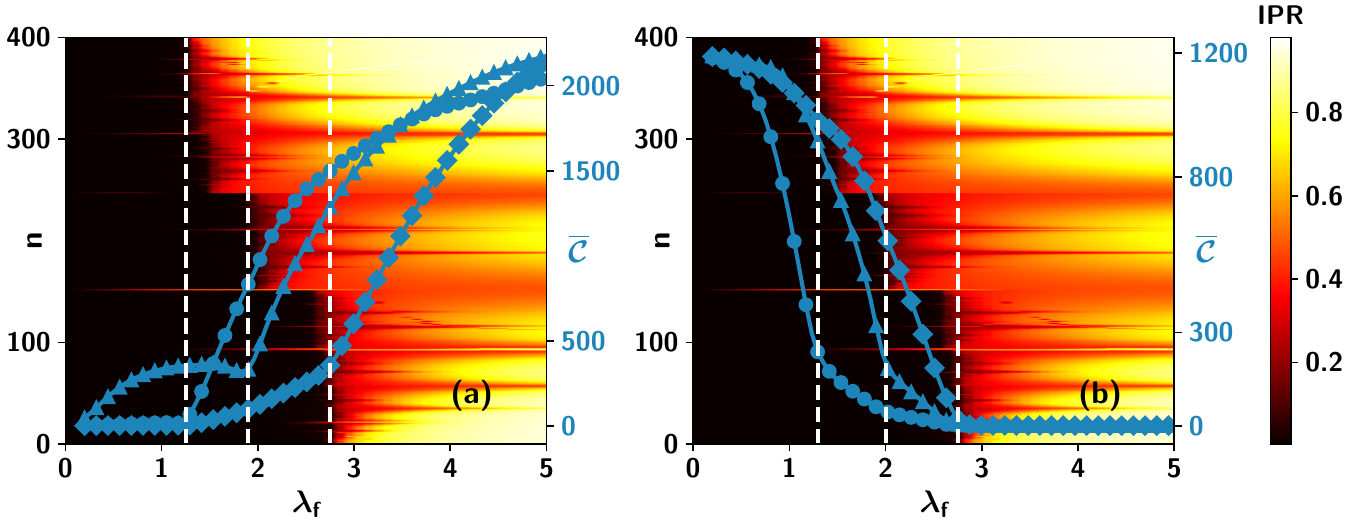}
    \caption{(a)\textbf{ Long-time averaged spread complexity for the forward and backward quench.} The inverse participation ratio (IPR) as a function of \(\lambda_f\) and the Krylov basis index \(n\) when \(\beta=0.3\) for the gAAH model where the initial states are the eigenstates of the Hamiltonian with \(\lambda_{in} =0\). The presence of a mobility edge is clear from the time-averaged spread complexity (values shown on the right side of the plot). The initial states are chosen to be the ground state (squares), a state from the middle spectrum (triangles), and the highest excited state (circles) for the forward quench. The three white vertical lines indicate the location of the nonanalyticities. (b) \textbf{Long-time averaged spread complexity for the backward quench.}  All other specifications are the same as in (a) except \(\lambda_{in}=100.0\) and the quenches being performed to the lower values of \(\lambda_f\). All axes are dimensionless.}
    \label{fig:beta0.3combined}
\end{figure}

We now move to the more general gAAH model, with \(\beta\ne 0\), where the localization-delocalization depends upon the energy of the state, giving rise to the mobility edge. More precisely, all the eigenstates with energy less than $E_0$ are delocalized while those with energy greater than $E_0$  are localized, where $E_0$, the mobility edge, is given by
\begin{equation} 
\beta E_0 = 2t - \lambda. 
\label{ME11} \end{equation}
This localization-delocalization transition can also be 
detected by IPR~\cite{Ganeshan2015} (see map plot in Fig. \ref{fig:beta0.3combined} for \(\beta =0.3\)). 

{\it Effectiveness of spread complexity for forward quench.} To capture this behavior via spread complexity, we study \(\overline{\mathcal{C}}\) for three different initial states of the gAAH model with \(\lambda_{in}=0\) and \(\beta\neq 0\), namely, (i) the ground state of the Hamiltonian, (ii) a state from the middle spectrum, and (iii) the highest excited state of the Hamiltonian. After the preparation of the initial state, the state is evolved by quenching the Hamiltonian with \(\lambda_f\). For all three types of initial states, \(\overline{\mathcal{C}}\) shows a kink at a particular value of \(\lambda_f\), whenever the initial energy crosses $E_0$, as given by Eq. (\ref{ME11}). The observations can be enumerated as follows:\\
(1) If the initial state is taken to be the ground state of the model, we observe that a sharp kink (nonanalyticity) in \(\overline{\mathcal{C}}\) appears around \(\lambda_f \approx 3\), exactly where the ground state shows a localization transition, as detected by IPR. After that, it increases monotonically with \(\lambda_f\) (see the dotted vertical line in Fig. \ref{fig:beta0.3combined}(a)). 

(2) For the highest excited state as the initial state, \(\overline{\mathcal{C}}\) remains zero till the transition point (\(\lambda_f \approx 1\)) and starts increasing monotonically with the increase of \(\lambda_f\).

(3) The different nonmonotonic behavior of \(\overline{\mathcal{C}}\) emerges for other initial states chosen from the middle of the spectrum.  Firstly, as for the other initial states, the nonanalyticity in \(\overline{\mathcal{C}}\) is observed at the transition point (\(\lambda_f \approx 2\)).  Secondly,  they behave nonmonotonically with \(\lambda_f \), especially near the transition point.  This nonmonotonic behavior cannot be seen when the initial states lie either in the low-energy sector or in the high-energy regime.  

{\it Note:}  These kinks become more and more pronounced as we go to higher system sizes. This shows that Krylov complexity can indeed determine the mobility edges~\footnote{We also notice that  the double derivative of \(\mathcal{C}\) with respect to \(\lambda_f\) exhibits the sharp kinks, thereby capable of capturing these mobility edges in the gAAH model.}.



{\it Backward quench.} Let us take the initial state of \(\hat{H}(\lambda_{in})\) with a high \(\lambda_{in}\), while \(\lambda_f\)s are chosen to be low. By choosing the ground, the middle, and the highest excited state of \(\lambda_{in}=100.0\) as initial states,  we calculate the time-averaged spread complexity (see Fig.~\ref{fig:beta0.3combined}(b)). The time-averaged spread complexity is almost vanishing when both \(\lambda_{in}\) and \(\lambda_f\) belong to the localized phase
while its nonanalytic behavior at the ME again emerges, depending on the initial states. Such observations establish that the time-averaged Krylov complexity not only captures the localization-delocalization transition, it is also sensitive to the presence of a mobility edge. 

\emph{The local density of states across phase boundaries.} Let us probe the origin of the observed behavior of the spread complexity. To do so, we analyze the local density of states (LDOS), which characterizes the distribution of energy eigenstates having finite overlap with a given initial state. The LDOS is defined as
\begin{eqnarray}
    \rho_{\psi_0}(E)=\sum_{r}|C_{\psi_0}^r|^2 \delta(E-E_{r}),
    \label{eq:ldos_def}
\end{eqnarray}
where $C_{\psi_0}^r=\langle\psi^{r}|\psi_0\rangle$ denotes the overlap between the initial state $|\psi_0\rangle$ and the eigenstate $|\psi^{r}\rangle$ of the post-quench Hamiltonian. In our calculations, the initial state $|\psi_0\rangle$ is chosen as the ground, states from the middle spectrum and the highest excited states of $\hat{H}(\lambda_{\mathrm{in}}=0)$, while $|\psi^{r}\rangle$ is an eigenstate of the quenched Hamiltonian $\hat{H}(\lambda_f)$ with eigenenergy $E_r$ (\(r=0,1, \ldots, N-1\)).

\begin{figure}  
    \centering \includegraphics[width=\linewidth]{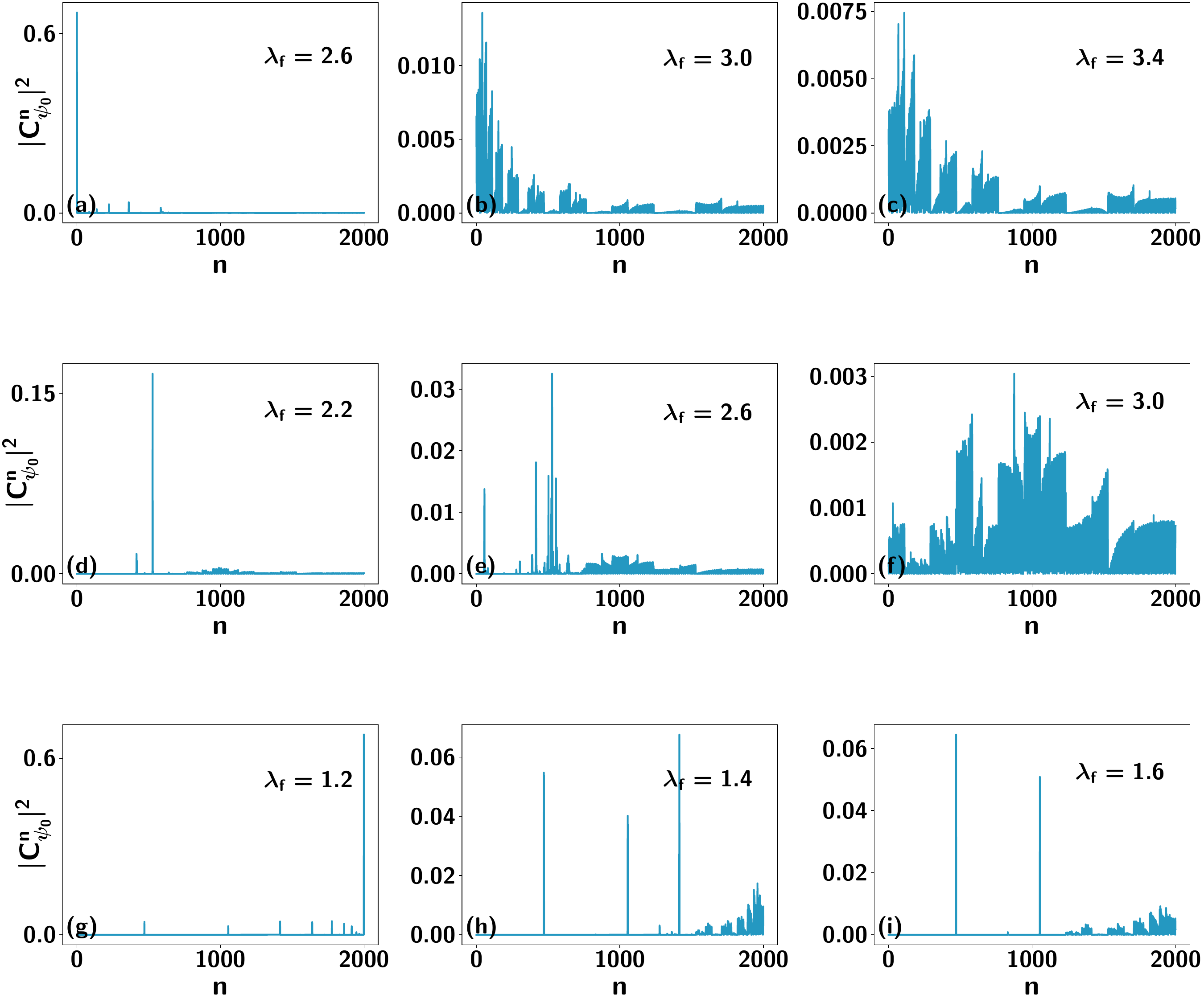}
    \caption{\textbf{Local density of states (LDOS) in the gAAH model}. (a)-(c) \(|C^{n}_{\psi_0}|^2\) as a function of \(n\), when \(|\psi_0\rangle\) is the ground state at \(\lambda=0\) and \(\lambda_f=2.6,3.0,3.4\) respectively. At \(\lambda=2.6\), \(|\psi_0\rangle\) has support on a very few eigenstates,  close to \(\lambda=3.0\) where the transition is; this distribution changes dramatically, and the support spreads from a few states to many of the eigenstates. This sudden spread gives rise to the kinks in the average value of complexity. For \(\lambda=3.4\), i.e., in the localized phase, the spread is even greater. (d)-(f) The initial state is the state from the middle spectrum of \(\lambda=0\) and \(\lambda_f=2.2,2.6,3.0\). (g)-(h) The initial state is the highest excited state at \(\lambda=0\) and \(\lambda_f=1.2,1.4,1.6\). All axes are dimensionless. }
    \label{fig:ldos_dist_beta0.3}
\end{figure}

For the AAH model, when $\lambda_f<2$ (extended phase), the initial state overlaps significantly with only a small subset of the eigenstates of the post-quench Hamiltonian. As a result, the LDOS is sharply peaked around a few eigenenergies, leading to relatively small values of the Lanczos coefficients. As $\lambda_f$ approaches, crosses the critical point $\lambda=2$ and moves to the localized phase, the LDOS broadens and spreads over a much larger portion of the spectrum. This broadening is directly reflected in the behavior of the Lanczos coefficients as depicted in Fig. \ref{fig:ldos_dist_beta0.3}. In particular, the first Lanczos coefficient is related to the variance of the LDOS through
\begin{eqnarray}
    b_1^2=\sum_{r}|C_{\psi_0}^r|^2 (E-\overline{E})^2=\sigma_{\psi_0}^2,
\end{eqnarray}
where $\overline{E}=\sum_{r}|C_{\psi_0}^r|^2 E_{r}$ is the mean of the LDOS and $\sigma_{\psi_0}^2$ is its variance. Consequently, the broadening of the LDOS across the localization transition results in an increase of the Lanczos coefficients, which, in turn, gives rise to the sharp enhancement observed in the long-time averaged spread complexity.

Let us elaborate: when the initial state is taken to be the ground state of the Hamiltonian in the extended phase before the mobility edge is crossed,   the LDOS is strongly concentrated around the ground-state energy (see Fig.~\ref{fig:ldos_dist_beta0.3}(a)). As the system approaches the localization-delocalization transition, the LDOS undergoes a substantial broadening (see Fig.~\ref{fig:ldos_dist_beta0.3}(b)), which is responsible for the kink observed in the long-time averaged spread complexity. Deep inside the localized regime, the LDOS remains broadly distributed over a large number of eigenstates. The same qualitative behavior is observed when the initial state is chosen from the middle of the spectrum or from the highest excited state.
Note that a similar analysis can be carried out for the backward quench by choosing the initial state from the localized phase.

\subsection{Analytical verification of results through moments in the limit of large quasi-periodic potential strength}
\label{sec:analytical_moments}

\begin{figure}  
    \centering \includegraphics[width=\linewidth]{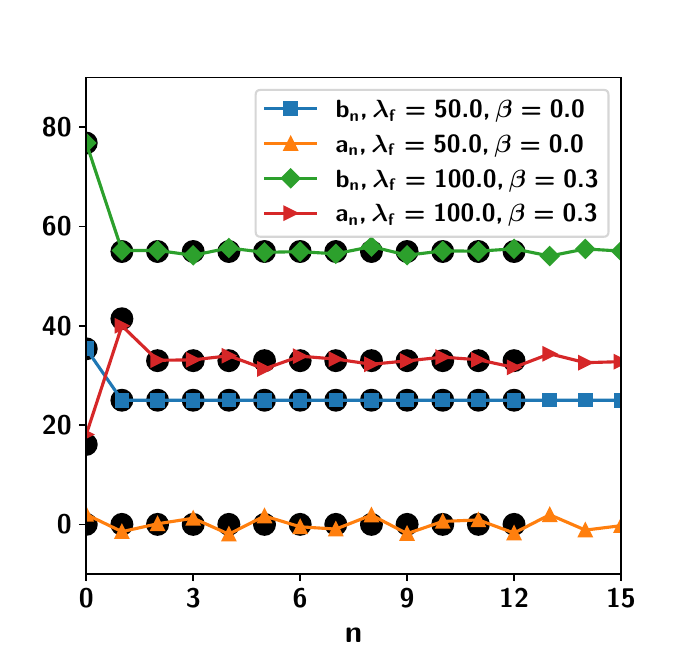}
    \caption{\textbf{The first few Lanczos coefficients are shown as a function of the Krylov basis index.} Triangles and squares represent the numerical results obtained using the Lanczos algorithm, whereas the black circles represent the analytical values calculated from the moments using the Markov-chain approach (with \(\beta =0\)). The initial state is taken to be the ground state at \(\lambda_{in}=0\). A similar study is carried out with \(\beta =0.3\).  All axes are dimensionless.}
    \label{fig:bnan_from_moments}
\end{figure}

Let us now analytically compute the expressions of Lanczos coefficients for a particular limit  for the generalized AAH model. In this case, the quenching operation is performed from \(\lambda_{in}=0\) to \(\lambda_f\to\infty\) and we determine the moments from the survival probability given in Eq.~(\ref{eq:moment_from_survival_probablity}). As in the case of \(\lambda_{in}=0\), the Hamiltonian consists of only  the hopping terms,  and hence the eigenstates can be written as
\begin{eqnarray}
|k\rangle = \frac{1}{\sqrt{N}} \sum_{j=1}^{N} e^{i k j a} \, c_j^\dagger |0\rangle,
\label{eq:bloch}
\end{eqnarray}
where \(|k\rangle\) denotes the state in the momentum space, with \(k = (2\pi/aN)\left( l - \frac{N}{2} \right), \quad l = 1,2,\dots,N\) and the energy of the state \(|k\rangle\) is given by \(-2 \cos (ka)\). In that representation,  the survival amplitude is given as
\begin{align}
S(t) 
&= \langle k | e^{-i H(\lambda_f) t} | k \rangle
= \sum_m e^{-i E_m(\lambda_f) t} 
\left| \langle \psi_m(\lambda_f) | k \rangle \right|^2,
\end{align}
where \(|\psi_m(\lambda_f)\rangle\) are the eigenstates of \(\hat{H}(\lambda_f)\). We note that in the case \(\lambda_f\to\infty\), the eigenstate of the Hamiltonian can be expressed as \(|\psi_m(\lambda_f \to \infty)\rangle 
= \sum_{j=1}^{N} \delta_{jm} \, c_j^\dagger |0\rangle\) with \(E_m=\lambda_f{\cos(2\pi q m)}/{(1-\beta \cos(2\pi q m)})\). Therefore, the expression for the survival probability reduces to
\begin{eqnarray}
S(t) &= \frac{1}{N} \sum_{m=1}^{N} 
\exp(-i \lambda_f\, t \frac{\cos\left( 2\pi q m \right)}{1-\beta \cos(2\pi q m)}) \nonumber \\ 
&\simeq \frac{1}{2\pi} \int_{-\pi}^{\pi} 
\exp\left(-i \lambda_f\, t \, \frac{\cos\theta}{1 - \beta \cos\theta} \right) \dd \theta,
\end{eqnarray}
where, in the last line, we use \(N\rightarrow \infty\) and apply Weyl's equidistribution theorem.
The moments can now be  computed by taking its \(n^{\text{th}}\) derivative of the survival probability which is given as
\begin{align}
\mu_n  &= \left. \frac{d^n}{dt^n} S(t) \right|_{t=0}  \nonumber \\ &
=\frac{(i \lambda_f)^n}{\beta^n} \Bigl[\sum_{k=0}^{n-1} 
\frac{(-1)^k \binom{n}{k}}{(1 - \beta^2)^{\frac{n-k}{2}}}\mathcal{P}_{n-k-1}\!\left( \frac{1}{\sqrt{1 - \beta^2}} \right)+1\Bigr],
\end{align} 
where \(\mathcal{P}_m(x)\) is the Legendre polynomial of the \(m\)-th order. From these moments, one can obtain the analytical expressions of the first Lanczos coefficients \(a_0\) and \(b_1\) which are given, respectively, as
\begin{eqnarray}
    a_0=\frac{\lambda_f}{\beta}\Bigl(\frac{1}{\sqrt{1-\beta^2}}+1\Bigr),
\end{eqnarray}
and 
\begin{eqnarray}
    b_1=\Bigg [ \frac{\lambda_f}{\beta}\Bigl(-\frac{(1+\frac{2\lambda_f}{\beta})}{\sqrt{1-\beta}}+(\frac{\lambda_f}{\beta}-1)+\frac{\lambda_f}{\beta}\frac{1}{(1-\beta^2)^{3/2}}\Bigr)\Bigg ]^{1/2}
\end{eqnarray}
For \(\beta=0\), the moments of the survival probability can be simplified more which is given as
\begin{eqnarray}
\text{If } n \text{ is odd:} \quad & \mu_n = 0, \nonumber \\
\text{If } n \text{ is even:} \quad 
& \mu_n = \frac{(i \lambda_f)^n}{\sqrt{\pi}}
\cdot \,
\frac{\Gamma\left(\frac{n+1}{2}\right)}
{\Gamma\left(\frac{n}{2}+1\right)},
\label{eq:moments_beta_zero}
\end{eqnarray}
where we notice that both \(S(t)\),  and \(\mu_n\) depend only upon the quenched parameter. Now, from the expression of \(\mu_n\), one can derive all the Lanczos coefficients by following the method of Motzkin paths, described in Ref.~\cite{Balasubramanian2022}. Now, for \(\beta=0\), we obtain that \(a_n\)s are zero, while \(b_n\)s can be found from \(\mu_n\)s. The first two \(b_n\)s read as \(b_1=\lambda_f/\sqrt{2}\) and \(b_2=\lambda_f/2\). A similar procedure can again be applied to obtain the analytical form of the survival probability for the backward quench, i.e., \(\lambda_{in} \to \infty\) to \(\lambda_f =0\). In this case, we obtain
\begin{align}
    S(t)
    &=\langle\psi_m|e^{-iH(\lambda_f)t}|\psi_m\rangle  \nonumber\\
    &=\frac{1}{N}\sum_{k} e^{i 2 \cos(ka)t} \nonumber\\&=\frac{a}{2 \pi} \int_{-\frac{\pi}{a}}^{\frac{\pi}{a}} e^{i2t \cos(ka)} dk=J_0(2t), 
\end{align}
by again taking \(N\rightarrow \infty\). 
The moments for backward quench are given by 
\begin{eqnarray}
\text{If } n \text{ is odd:} \quad & \mu_n = 0, \nonumber \\
\text{If } n \text{ is even:} \quad 
& \mu_n = \frac{(2i)^n}{\sqrt{\pi}}
\cdot \,
\frac{\Gamma\left(\frac{n+1}{2}\right)}
{\Gamma\left(\frac{n}{2}+1\right)}.
\label{eq:moments_backward_gaah}
\end{eqnarray}
Hence the \(a_n\)s are also zero here and \(b_1=\sqrt{2}\) and \(b_2=1\). In order to verify the Lanczos coefficients numerically, in Fig.~\ref{fig:bnan_from_moments}, we plot the Lanczos coefficients up to \(n=15\) which shows that the numerical results match the values derived from the analytical expressions of moments.

\section{Effectiveness of spread complexity beyond NN hoping}
\label{sec:long_range_spread_complexity}

Till now, we have investigated the behavior of spread complexity in the generalized AAH model with nearest-neighbor hopping. We now extend our analysis beyond NN hopping to determine whether spread complexity remains an equally effective probe of localization and dynamical phase transitions~\cite{Heyl2018} in this more general setting. In order to start the analysis, we first consider next-nearest neighbor (NNN) hopping.

\begin{figure}  
    \centering \includegraphics[width=\linewidth]{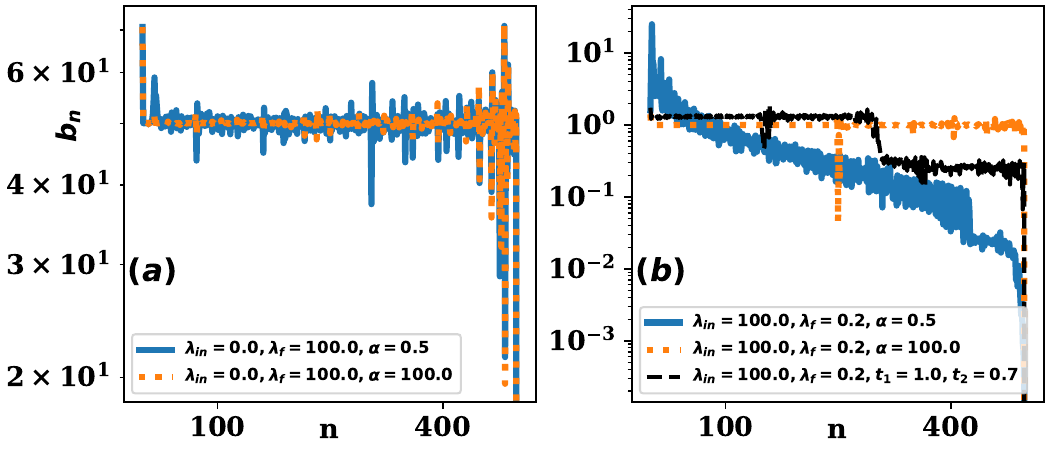}
    \caption{\textbf{Comparison of the Lanczos coefficients \(b_n\) for the long-range, NNN  and NN AAH model.} (a) Variation of \(b_n\) with \(n\) for the forward quench of the Hamiltonian with the NN (\(\alpha=100\)) and LR  hopping \(\alpha=0.5\). Their behavior is nearly identical. (b) Comparison of \(b_n\)s vs \(n\) for the backward quench. In the nearest neighbor case, \(b_n\) remains close \(1\) for all values of \(n\) while for the NNN hopping, it is constant before dropping sharply near \(n=D_K\). In contrast, for long-range hopping, \(b_n\) decreases continuously with \(n\).  All axes are dimensionless.}
    \label{fig:bn_longrange_shortrange}
\end{figure}

\emph{Spread complexity in the presence of next-nearest neighbor (NNN) hopping along with NN one.} The Hamiltonian of the AAH model having both NN and NNN hoppings can be represented as 
\begin{align}
    \hat{H}_{NNN} 
    &= t_1\sum (c_j^\dagger c_{j+1}+ \mathrm{h.c.})+t_2\sum (c_j^\dagger c_{j+2}+ \mathrm{h.c.})\nonumber \\
    &+\lambda \sum_{i} \cos(2\pi q i + \phi) c_i^\dagger c_{i},
\end{align}
where \(t_{1(2)}\) is the strength of NN (NNN) hopping. The introduction of NNN hopping breaks the self-duality and creates a mobility edge although the eigenstates at \(\lambda=0\) do not change, but the eigenvalues change, which are given by 
\begin{eqnarray}
    E_k=-2t_1 \cos (ka)-2t_2 \cos(2ka).
\end{eqnarray}
Since the eigenstates at \(\lambda=0\) remain the same, the survival probability and the moments also remain the same as Eq.~ (\ref{eq:moments_beta_zero}), and as a result, the Lanczos coefficients remain unchanged for the case of forward quench. Note, however, that the Lanczos coefficients are different for the backward quench as the eigenstates of the system are not the same as the NN Hamiltonian. For the backward quench, the return amplitude can be written as 
\begin{eqnarray}
    G_k(t)=\frac{1}{N}\sum_{k} \exp \big[i(2t_1\cos(ka)+2t_2\cos(2ka))t\big],
\end{eqnarray}
and when \(N\rightarrow\infty\), we obtain
\begin{eqnarray}
    \nonumber G_k(t)=\frac{a}{2\pi}\int_{-\frac{\pi}{a}}^{\frac{\pi}{a}} \exp\big [{i(2t_1\cos(ka)+2t_2\cos(2ka))t}\big] dk,\\
\end{eqnarray}
which leads to the moments
\begin{align}
\mu_n
&=
\frac{i^n}{2\pi}
\int_{-\pi}^{\pi}
\Bigl(2t_1\cos(k)+2t_2\cos(2k)\Bigr)^n\,dk \nonumber
\\&=
\frac{i^n\,2^n}{\pi}
\sum_{\substack{p=0\\ p\ \mathrm{even}}}^{n}
\sum_{q=0}^{n-p}
\binom{n}{p}
\binom{n-p}{q}
t_1^p t_2^{\,n-p} 
2^q
(-1)^{\,n-p-q} \nonumber\\
&\qquad\qquad{\sqrt{\pi}\,
\Gamma\!\left(\dfrac{p+2q+1}{2}\right)}\Big /
{\Gamma\!\left(\dfrac{p+2q}{2}+1\right)}.
\end{align}
Now, from \(\mu_n\), the Lanczos coefficients can, in principle, be derived for the backward quench, especially when \(n\) is small. We numerically compute \(b_n\)s in Fig.~\ref{fig:bn_longrange_shortrange} by the black dots. We find that, similar to the NN case, \(b_n\) remains constant with the initial increase of \(n\), but an abrupt change of \(b_n\) is observed for large \(n\), which clearly deviates from the NN case. It indicates that introducing hopping beyond NN ones may lead to some interesting phenomena which are typically absent in the NN case. 


\emph{Long-range AAH model.} Let us now consider the long-range AAH model described in Eq.~(\ref{eq:ham}) with $\beta=0$, where long-range hopping alone gives rise to a mobility edge even in the absence of the deformation parameter $\beta$. In this case, the energy-dependent mobility edge is given as~\cite{Biddle2011}
\begin{equation}
    E=\lambda\cosh(\alpha\ln2)-t,
\end{equation}
which now depends on the range of hopping, \(\alpha\). 

\begin{figure}  
    \centering \includegraphics[width=\linewidth]{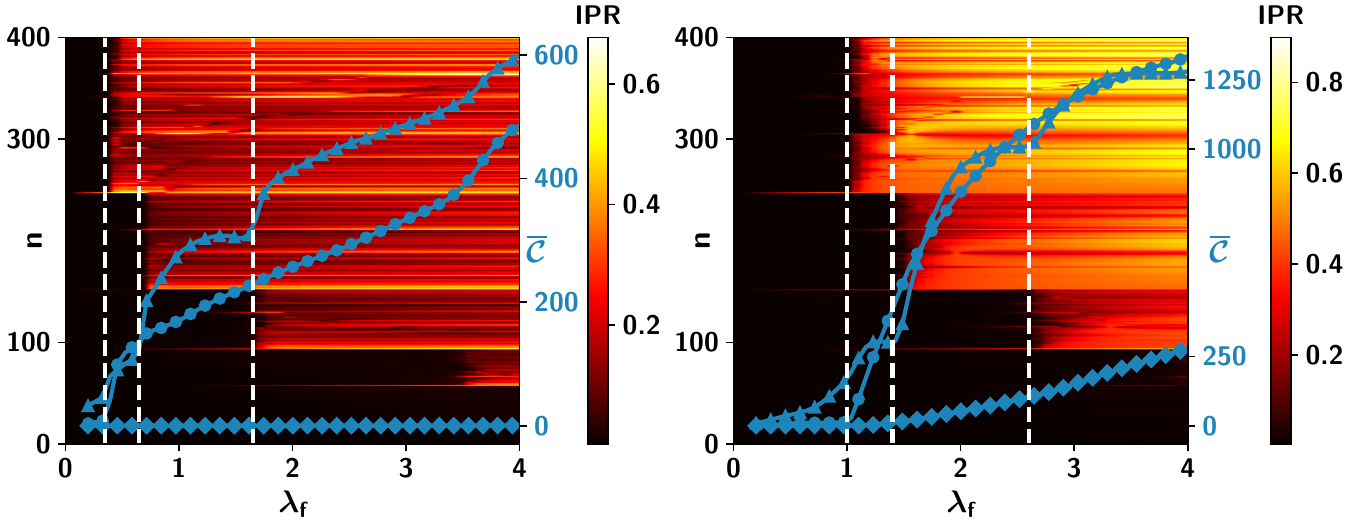}
    \caption{\textbf{ Map plot of IPR as a function of \(\lambda_f\) and the eigenstate index is plotted when long-range hopping is present for the forward quench with (a) \(\alpha=0.5\) (b) \(\alpha=1.5\).} The long time-averaged spread complexity for three quenches of the ground state (diamonds), a state from the middle spectrum (upper triangles), and the highest excited state (circles) as initial states are plotted on the right axis of the plot. The spread complexity shows a kink whenever a mobility edge is crossed.  All axes are dimensionless.}
    \label{fig:lr_0.5_1.5}
\end{figure}

To verify the existence of the mobility edge, we compute the inverse participation ratio (IPR). As shown in the map plot of Fig.~\ref{fig:lr_0.5_1.5} for $\alpha=0.5$ and \(\alpha =1.5\), the IPR clearly reveals the presence of a mobility edge. Similar to the generalized AAH model, the localization properties of the eigenstates are strongly energy-dependent, even when $\beta=0$. We also note that, unlike the AAH model, the IPR does not approach unity in the localized regime, since the eigenstates above the mobility edge are multifractal rather than exponentially localized for low \(\alpha\).

We next investigate the behavior of the spread complexity for the long-range model. As in the NN case, we consider three representative initial states prepared at $\lambda_{in}=0$: the ground state, an eigenstate from the middle of the spectrum, and the highest excited state. Each of these states is evolved after a sudden quench to different values of $\lambda_f$, and the corresponding long-time averaged spread complexity, $\overline{\mathcal{C}}$, is computed, as shown in Fig.~\ref{fig:lr_0.5_1.5}.

Our results demonstrate that $\overline{\mathcal{C}}$ remains an efficient probe of the mobility edge even in the presence of long-range hopping. For every choice of the initial state, $\overline{\mathcal{C}}$ exhibits a pronounced kink, whenever the post-quench parameters cross the mobility edge associated with the energy of the initial state. In particular, when the initial state is the ground state, $\overline{\mathcal{C}}$ remains nearly zero throughout the regime $\lambda_f<4$. In contrast, when the highest excited state is chosen as the initial state, $\overline{\mathcal{C}}$ remains vanishing up to the corresponding critical value of $\lambda_f$ and starts increasing once the mobility edge is crossed. For initial states chosen from the middle of the spectrum, $\overline{\mathcal{C}}$ displays a non-monotonic dependence on $\lambda_f$, with distinct kinks marking the locations of the mobility edges, closely resembling the behavior observed in the generalized nearest-neighbor AAH model.

We now turn to the behavior of the Lanczos coefficients $b_n$. For the forward quench protocol, we observe that $b_n$ decreases sharply from $n=0$ to $n=1$ and subsequently fluctuates around an approximately constant value, as illustrated in Fig.~\ref{fig:bn_longrange_shortrange}. This behavior closely resembles that of the NN model, although the fluctuations are noticeably larger in the long-range case. Moreover, the average value of $b_n$ is found to be nearly identical to that of the NN model. This can be understood from the fact that the eigenstates at $\lambda=0$ are identical to those of the NN AAH model described by Eq.~(\ref{eq:bloch}); only the energy spectrum is modified according to \(E_k=-2t\sum_{j=1}^{L/2}\frac{\cos(jk)}{j^\alpha}\). Consequently, the average Lanczos coefficients remain almost unchanged because the survival amplitude is primarily determined by the structure of the initial and final eigenstates, which is preserved for LR hopping. The enhanced fluctuations observed in the long-range model are due to the multifractal nature of the eigenstates.


 {\it Backward quench in long-range system.} In the case of the backward quench, the moments of the survival amplitude can be evaluated analytically in the thermodynamic limit ($N\rightarrow\infty$) and are given by
\begin{align}
\mu_n &= \frac{i^n}{2\pi} \int_{-\pi}^{\pi}
\left(2 \sum_{j=1}^{L/2} \frac{\cos(jka)}{j^{\alpha}}\right)^n
\,dk.
\end{align}
Again, using these moments, the Lanczos coefficients may be obtained recursively for small \(n\). See Fig.~\ref{fig:bn_longrange_shortrange}(b) for numerical computation of these \(b_n\)s. Interestingly, in contrast to the AAH case, we find that the Lanczos coefficients $b_n$ decay gradually with increasing Krylov index $n$ and eventually approach zero. This behavior closely resembles that observed in non-integrable quantum spin chains~\cite{Bhattacharya_prd_2024}. Our results therefore suggest that the long-range hopping qualitatively modifies the Krylov dynamics, driving the behavior of the Lanczos coefficients towards that typically associated with non-integrable systems as the range of hopping increases.

\section{Conclusion}
\label{sec:conclu}

Krylov complexity, referred to as spread complexity in the state space, quantifies the spreading of a quantum state over the Krylov basis generated by the Hamiltonian governing its dynamics. A key property of the Krylov basis is that it minimizes the spread of the evolving state, making spread complexity a natural measure for characterizing quantum dynamics. Consequently, it has emerged as a promising diagnostic for identifying different equilibrium phases of quantum Hamiltonians through dynamics.

We employed the spread complexity here to investigate localization-- delocalization transitions in the long-range (LR) Aubry--Andr\'e--Harper (AAH) model, a paradigmatic quasiperiodic system 
that hosts extended, localized, and mobility edges. We showed that the long-time averaged spread complexity serves as an effective order parameter for distinguishing the extended and localized phases in the nearest-neighbor (NN) Aubry-Andr\'e model. For the generalized AAH model with mobility edges, we found that long-time averaged spread complexity 
develops pronounced nonanlytic features whenever the post-quench parameters cross a mobility edge, as obtained through the inverse participation ratio (IPR).  For the long-range AAH model with power-law hopping, the IPR does not approach unity in the localized phase, in contrast to the NN generalized AAH model. Nevertheless, the long-time-averaged spread complexity remains sensitive to the localization properties of the system. In particular, it accurately identifies the mobility edge for initial states spanning different energy sectors, demonstrating its robustness as a dynamical probe even in the presence of long-range hoppings. We also demonstrated that the Lanczos coefficients can be obtained analytically from the moments of the survival amplitude when the initial and post-quench Hamiltonians are chosen at the extreme points of the phase diagram, for both the short-range and LR models, providing additional insight into the underlying dynamics. Our results establish spread complexity as a simple and efficient means for identifying different phases and accurately locating mobility edges in single-particle quasiperiodic systems. We anticipate that these findings can motivate further investigations of Krylov complexity in a broader class of disordered and quasiperiodic quantum systems.



\acknowledgements
We acknowledge support from the project entitled ``Technology Vertical - Quantum Communication'' under the National Quantum Mission of the Department of Science and Technology (DST)  (Sanction Order No. DST/QTC/NQM/QComm/$2024/2$ (G)). This research was carried out and financed within the framework of the second Swiss Contribution MAPS (Grant No. 230870).

\bibliography{ref.bib}

\begin{thebibliography}{76}%
\makeatletter
\providecommand \@ifxundefined [1]{%
 \@ifx{#1\undefined}
}%
\providecommand \@ifnum [1]{%
 \ifnum #1\expandafter \@firstoftwo
 \else \expandafter \@secondoftwo
 \fi
}%
\providecommand \@ifx [1]{%
 \ifx #1\expandafter \@firstoftwo
 \else \expandafter \@secondoftwo
 \fi
}%
\providecommand \natexlab [1]{#1}%
\providecommand \enquote  [1]{``#1''}%
\providecommand \bibnamefont  [1]{#1}%
\providecommand \bibfnamefont [1]{#1}%
\providecommand \citenamefont [1]{#1}%
\providecommand \href@noop [0]{\@secondoftwo}%
\providecommand \href [0]{\begingroup \@sanitize@url \@href}%
\providecommand \@href[1]{\@@startlink{#1}\@@href}%
\providecommand \@@href[1]{\endgroup#1\@@endlink}%
\providecommand \@sanitize@url [0]{\catcode `\\12\catcode `\$12\catcode
  `\&12\catcode `\#12\catcode `\^12\catcode `\_12\catcode `\%12\relax}%
\providecommand \@@startlink[1]{}%
\providecommand \@@endlink[0]{}%
\providecommand \url  [0]{\begingroup\@sanitize@url \@url }%
\providecommand \@url [1]{\endgroup\@href {#1}{\urlprefix }}%
\providecommand \urlprefix  [0]{URL }%
\providecommand \Eprint [0]{\href }%
\providecommand \doibase [0]{https://doi.org/}%
\providecommand \selectlanguage [0]{\@gobble}%
\providecommand \bibinfo  [0]{\@secondoftwo}%
\providecommand \bibfield  [0]{\@secondoftwo}%
\providecommand \translation [1]{[#1]}%
\providecommand \BibitemOpen [0]{}%
\providecommand \bibitemStop [0]{}%
\providecommand \bibitemNoStop [0]{.\EOS\space}%
\providecommand \EOS [0]{\spacefactor3000\relax}%
\providecommand \BibitemShut  [1]{\csname bibitem#1\endcsname}%
\let\auto@bib@innerbib\@empty
\bibitem [{\citenamefont {Nielsen}(2006)}]{Nielsen2006}%
  \BibitemOpen
  \bibfield  {author} {\bibinfo {author} {\bibfnamefont {M.}~\bibnamefont
  {Nielsen}},\ }\bibfield  {title} {\bibinfo {title} {A geometric approach to
  quantum circuit lower bounds},\ }\href {https://doi.org/10.26421/qic6.3-2}
  {\bibfield  {journal} {\bibinfo  {journal} {Quantum Information and
  Computation}\ }\textbf {\bibinfo {volume} {6}},\ \bibinfo {pages} {213–262}
  (\bibinfo {year} {2006})}\BibitemShut {NoStop}%
\bibitem [{\citenamefont {Dowling}\ and\ \citenamefont
  {Nielsen}(2008)}]{Dowling2008}%
  \BibitemOpen
  \bibfield  {author} {\bibinfo {author} {\bibfnamefont {M.}~\bibnamefont
  {Dowling}}\ and\ \bibinfo {author} {\bibfnamefont {M.}~\bibnamefont
  {Nielsen}},\ }\bibfield  {title} {\bibinfo {title} {The geometry of quantum
  computation},\ }\href {https://doi.org/10.26421/qic8.10-1} {\bibfield
  {journal} {\bibinfo  {journal} {Quantum Information and Computation}\
  }\textbf {\bibinfo {volume} {8}},\ \bibinfo {pages} {861–899} (\bibinfo
  {year} {2008})}\BibitemShut {NoStop}%
\bibitem [{\citenamefont {Kolmogorov}(1998)}]{Kolmogorov1998}%
  \BibitemOpen
  \bibfield  {author} {\bibinfo {author} {\bibfnamefont {A.}~\bibnamefont
  {Kolmogorov}},\ }\bibfield  {title} {\bibinfo {title} {On tables of random
  numbers},\ }\href {https://doi.org/10.1016/s0304-3975(98)00075-9} {\bibfield
  {journal} {\bibinfo  {journal} {Theoretical Computer Science}\ }\textbf
  {\bibinfo {volume} {207}},\ \bibinfo {pages} {387–395} (\bibinfo {year}
  {1998})}\BibitemShut {NoStop}%
\bibitem [{\citenamefont {Leone}\ \emph {et~al.}(2022)\citenamefont {Leone},
  \citenamefont {Oliviero},\ and\ \citenamefont {Hamma}}]{leone_prl_2022}%
  \BibitemOpen
  \bibfield  {author} {\bibinfo {author} {\bibfnamefont {L.}~\bibnamefont
  {Leone}}, \bibinfo {author} {\bibfnamefont {S.~F.~E.}\ \bibnamefont
  {Oliviero}},\ and\ \bibinfo {author} {\bibfnamefont {A.}~\bibnamefont
  {Hamma}},\ }\bibfield  {title} {\bibinfo {title} {Stabilizer r\'enyi
  entropy},\ }\href {https://doi.org/10.1103/PhysRevLett.128.050402} {\bibfield
   {journal} {\bibinfo  {journal} {Phys. Rev. Lett.}\ }\textbf {\bibinfo
  {volume} {128}},\ \bibinfo {pages} {050402} (\bibinfo {year}
  {2022})}\BibitemShut {NoStop}%
\bibitem [{\citenamefont {Nandy}\ \emph
  {et~al.}(2025{\natexlab{a}})\citenamefont {Nandy}, \citenamefont
  {Matsoukas-Roubeas}, \citenamefont {Martínez-Azcona}, \citenamefont
  {Dymarsky},\ and\ \citenamefont {del Campo}}]{Nandy2025}%
  \BibitemOpen
  \bibfield  {author} {\bibinfo {author} {\bibfnamefont {P.}~\bibnamefont
  {Nandy}}, \bibinfo {author} {\bibfnamefont {A.~S.}\ \bibnamefont
  {Matsoukas-Roubeas}}, \bibinfo {author} {\bibfnamefont {P.}~\bibnamefont
  {Martínez-Azcona}}, \bibinfo {author} {\bibfnamefont {A.}~\bibnamefont
  {Dymarsky}},\ and\ \bibinfo {author} {\bibfnamefont {A.}~\bibnamefont {del
  Campo}},\ }\bibfield  {title} {\bibinfo {title} {Quantum dynamics in krylov
  space: Methods and applications},\ }\href
  {https://doi.org/10.1016/j.physrep.2025.05.001} {\bibfield  {journal}
  {\bibinfo  {journal} {Physics Reports}\ }\textbf {\bibinfo {volume}
  {1125-1128}},\ \bibinfo {pages} {1–82} (\bibinfo {year}
  {2025}{\natexlab{a}})}\BibitemShut {NoStop}%
\bibitem [{\citenamefont {Parker}\ \emph {et~al.}(2019)\citenamefont {Parker},
  \citenamefont {Cao}, \citenamefont {Avdoshkin}, \citenamefont {Scaffidi},\
  and\ \citenamefont {Altman}}]{Parker2019}%
  \BibitemOpen
  \bibfield  {author} {\bibinfo {author} {\bibfnamefont {D.~E.}\ \bibnamefont
  {Parker}}, \bibinfo {author} {\bibfnamefont {X.}~\bibnamefont {Cao}},
  \bibinfo {author} {\bibfnamefont {A.}~\bibnamefont {Avdoshkin}}, \bibinfo
  {author} {\bibfnamefont {T.}~\bibnamefont {Scaffidi}},\ and\ \bibinfo
  {author} {\bibfnamefont {E.}~\bibnamefont {Altman}},\ }\bibfield  {title}
  {\bibinfo {title} {A universal operator growth hypothesis},\ }\bibfield
  {journal} {\bibinfo  {journal} {Physical Review X}\ }\textbf {\bibinfo
  {volume} {9}},\ \href {https://doi.org/10.1103/physrevx.9.041017}
  {10.1103/physrevx.9.041017} (\bibinfo {year} {2019})\BibitemShut {NoStop}%
\bibitem [{\citenamefont {Avdoshkin}\ and\ \citenamefont
  {Dymarsky}(2020)}]{Avdoshkin2020}%
  \BibitemOpen
  \bibfield  {author} {\bibinfo {author} {\bibfnamefont {A.}~\bibnamefont
  {Avdoshkin}}\ and\ \bibinfo {author} {\bibfnamefont {A.}~\bibnamefont
  {Dymarsky}},\ }\bibfield  {title} {\bibinfo {title} {Euclidean operator
  growth and quantum chaos},\ }\bibfield  {journal} {\bibinfo  {journal}
  {Physical Review Research}\ }\textbf {\bibinfo {volume} {2}},\ \href
  {https://doi.org/10.1103/physrevresearch.2.043234}
  {10.1103/physrevresearch.2.043234} (\bibinfo {year} {2020})\BibitemShut
  {NoStop}%
\bibitem [{\citenamefont {Barbón}\ \emph {et~al.}(2019)\citenamefont
  {Barbón}, \citenamefont {Rabinovici}, \citenamefont {Shir},\ and\
  \citenamefont {Sinha}}]{Barbn2019}%
  \BibitemOpen
  \bibfield  {author} {\bibinfo {author} {\bibfnamefont {J.}~\bibnamefont
  {Barbón}}, \bibinfo {author} {\bibfnamefont {E.}~\bibnamefont {Rabinovici}},
  \bibinfo {author} {\bibfnamefont {R.}~\bibnamefont {Shir}},\ and\ \bibinfo
  {author} {\bibfnamefont {R.}~\bibnamefont {Sinha}},\ }\bibfield  {title}
  {\bibinfo {title} {On the evolution of operator complexity beyond
  scrambling},\ }\bibfield  {journal} {\bibinfo  {journal} {Journal of High
  Energy Physics}\ }\textbf {\bibinfo {volume} {2019}},\ \href
  {https://doi.org/10.1007/jhep10(2019)264} {10.1007/jhep10(2019)264} (\bibinfo
  {year} {2019})\BibitemShut {NoStop}%
\bibitem [{\citenamefont {Rabinovici}\ \emph {et~al.}(2021)\citenamefont
  {Rabinovici}, \citenamefont {Sánchez-Garrido}, \citenamefont {Shir},\ and\
  \citenamefont {Sonner}}]{Rabinovici2021}%
  \BibitemOpen
  \bibfield  {author} {\bibinfo {author} {\bibfnamefont {E.}~\bibnamefont
  {Rabinovici}}, \bibinfo {author} {\bibfnamefont {A.}~\bibnamefont
  {Sánchez-Garrido}}, \bibinfo {author} {\bibfnamefont {R.}~\bibnamefont
  {Shir}},\ and\ \bibinfo {author} {\bibfnamefont {J.}~\bibnamefont {Sonner}},\
  }\bibfield  {title} {\bibinfo {title} {Operator complexity: a journey to the
  edge of krylov space},\ }\bibfield  {journal} {\bibinfo  {journal} {Journal
  of High Energy Physics}\ }\textbf {\bibinfo {volume} {2021}},\ \href
  {https://doi.org/10.1007/jhep06(2021)062} {10.1007/jhep06(2021)062} (\bibinfo
  {year} {2021})\BibitemShut {NoStop}%
\bibitem [{\citenamefont {Rabinovici}\ \emph
  {et~al.}(2022{\natexlab{a}})\citenamefont {Rabinovici}, \citenamefont
  {Sánchez-Garrido}, \citenamefont {Shir},\ and\ \citenamefont
  {Sonner}}]{Rabinovici2022_a}%
  \BibitemOpen
  \bibfield  {author} {\bibinfo {author} {\bibfnamefont {E.}~\bibnamefont
  {Rabinovici}}, \bibinfo {author} {\bibfnamefont {A.}~\bibnamefont
  {Sánchez-Garrido}}, \bibinfo {author} {\bibfnamefont {R.}~\bibnamefont
  {Shir}},\ and\ \bibinfo {author} {\bibfnamefont {J.}~\bibnamefont {Sonner}},\
  }\bibfield  {title} {\bibinfo {title} {Krylov complexity from integrability
  to chaos},\ }\bibfield  {journal} {\bibinfo  {journal} {Journal of High
  Energy Physics}\ }\textbf {\bibinfo {volume} {2022}},\ \href
  {https://doi.org/10.1007/jhep07(2022)151} {10.1007/jhep07(2022)151} (\bibinfo
  {year} {2022}{\natexlab{a}})\BibitemShut {NoStop}%
\bibitem [{\citenamefont {Rabinovici}\ \emph
  {et~al.}(2022{\natexlab{b}})\citenamefont {Rabinovici}, \citenamefont
  {Sánchez-Garrido}, \citenamefont {Shir},\ and\ \citenamefont
  {Sonner}}]{Rabinovici2022_b}%
  \BibitemOpen
  \bibfield  {author} {\bibinfo {author} {\bibfnamefont {E.}~\bibnamefont
  {Rabinovici}}, \bibinfo {author} {\bibfnamefont {A.}~\bibnamefont
  {Sánchez-Garrido}}, \bibinfo {author} {\bibfnamefont {R.}~\bibnamefont
  {Shir}},\ and\ \bibinfo {author} {\bibfnamefont {J.}~\bibnamefont {Sonner}},\
  }\bibfield  {title} {\bibinfo {title} {Krylov localization and suppression of
  complexity},\ }\bibfield  {journal} {\bibinfo  {journal} {Journal of High
  Energy Physics}\ }\textbf {\bibinfo {volume} {2022}},\ \href
  {https://doi.org/10.1007/jhep03(2022)211} {10.1007/jhep03(2022)211} (\bibinfo
  {year} {2022}{\natexlab{b}})\BibitemShut {NoStop}%
\bibitem [{\citenamefont {Jian}\ \emph {et~al.}(2021)\citenamefont {Jian},
  \citenamefont {Swingle},\ and\ \citenamefont {Xian}}]{Jian2021}%
  \BibitemOpen
  \bibfield  {author} {\bibinfo {author} {\bibfnamefont {S.-K.}\ \bibnamefont
  {Jian}}, \bibinfo {author} {\bibfnamefont {B.}~\bibnamefont {Swingle}},\ and\
  \bibinfo {author} {\bibfnamefont {Z.-Y.}\ \bibnamefont {Xian}},\ }\bibfield
  {title} {\bibinfo {title} {Complexity growth of operators in the syk model
  and in jt gravity},\ }\bibfield  {journal} {\bibinfo  {journal} {Journal of
  High Energy Physics}\ }\textbf {\bibinfo {volume} {2021}},\ \href
  {https://doi.org/10.1007/jhep03(2021)014} {10.1007/jhep03(2021)014} (\bibinfo
  {year} {2021})\BibitemShut {NoStop}%
\bibitem [{\citenamefont {Menzler}\ and\ \citenamefont
  {Jha}(2024{\natexlab{a}})}]{Menzler_prb_2024}%
  \BibitemOpen
  \bibfield  {author} {\bibinfo {author} {\bibfnamefont {H.~G.}\ \bibnamefont
  {Menzler}}\ and\ \bibinfo {author} {\bibfnamefont {R.}~\bibnamefont {Jha}},\
  }\bibfield  {title} {\bibinfo {title} {Krylov delocalization/localization
  across ergodicity breaking},\ }\href
  {https://doi.org/10.1103/PhysRevB.110.125137} {\bibfield  {journal} {\bibinfo
   {journal} {Phys. Rev. B}\ }\textbf {\bibinfo {volume} {110}},\ \bibinfo
  {pages} {125137} (\bibinfo {year} {2024}{\natexlab{a}})}\BibitemShut
  {NoStop}%
\bibitem [{\citenamefont {Dymarsky}\ and\ \citenamefont
  {Smolkin}(2021)}]{Dymarsky2021}%
  \BibitemOpen
  \bibfield  {author} {\bibinfo {author} {\bibfnamefont {A.}~\bibnamefont
  {Dymarsky}}\ and\ \bibinfo {author} {\bibfnamefont {M.}~\bibnamefont
  {Smolkin}},\ }\bibfield  {title} {\bibinfo {title} {Krylov complexity in
  conformal field theory},\ }\href
  {https://doi.org/10.1103/PhysRevD.104.L081702} {\bibfield  {journal}
  {\bibinfo  {journal} {Phys. Rev. D}\ }\textbf {\bibinfo {volume} {104}},\
  \bibinfo {pages} {L081702} (\bibinfo {year} {2021})}\BibitemShut {NoStop}%
\bibitem [{\citenamefont {Nizami}\ and\ \citenamefont
  {Shrestha}(2024)}]{nizami_pre_2024}%
  \BibitemOpen
  \bibfield  {author} {\bibinfo {author} {\bibfnamefont {A.~A.}\ \bibnamefont
  {Nizami}}\ and\ \bibinfo {author} {\bibfnamefont {A.~W.}\ \bibnamefont
  {Shrestha}},\ }\bibfield  {title} {\bibinfo {title} {Spread complexity and
  quantum chaos for periodically driven spin chains},\ }\href
  {https://doi.org/10.1103/PhysRevE.110.034201} {\bibfield  {journal} {\bibinfo
   {journal} {Phys. Rev. E}\ }\textbf {\bibinfo {volume} {110}},\ \bibinfo
  {pages} {034201} (\bibinfo {year} {2024})}\BibitemShut {NoStop}%
\bibitem [{\citenamefont {Suchsland}\ \emph {et~al.}(2025)\citenamefont
  {Suchsland}, \citenamefont {Moessner},\ and\ \citenamefont
  {Claeys}}]{Suchsland2025}%
  \BibitemOpen
  \bibfield  {author} {\bibinfo {author} {\bibfnamefont {P.}~\bibnamefont
  {Suchsland}}, \bibinfo {author} {\bibfnamefont {R.}~\bibnamefont
  {Moessner}},\ and\ \bibinfo {author} {\bibfnamefont {P.~W.}\ \bibnamefont
  {Claeys}},\ }\bibfield  {title} {\bibinfo {title} {Krylov complexity and
  trotter transitions in unitary circuit dynamics},\ }\href
  {https://doi.org/10.1103/PhysRevB.111.014309} {\bibfield  {journal} {\bibinfo
   {journal} {Phys. Rev. B}\ }\textbf {\bibinfo {volume} {111}},\ \bibinfo
  {pages} {014309} (\bibinfo {year} {2025})}\BibitemShut {NoStop}%
\bibitem [{\citenamefont {Kolganov}\ and\ \citenamefont
  {Trunin}(2025)}]{Streamlined_2025}%
  \BibitemOpen
  \bibfield  {author} {\bibinfo {author} {\bibfnamefont {N.}~\bibnamefont
  {Kolganov}}\ and\ \bibinfo {author} {\bibfnamefont {D.~A.}\ \bibnamefont
  {Trunin}},\ }\bibfield  {title} {\bibinfo {title} {Streamlined krylov
  construction and classification of ergodic floquet systems},\ }\href
  {https://doi.org/10.1103/PhysRevE.111.L052202} {\bibfield  {journal}
  {\bibinfo  {journal} {Phys. Rev. E}\ }\textbf {\bibinfo {volume} {111}},\
  \bibinfo {pages} {L052202} (\bibinfo {year} {2025})}\BibitemShut {NoStop}%
\bibitem [{\citenamefont {Bhattacharya}\ \emph {et~al.}(2022)\citenamefont
  {Bhattacharya}, \citenamefont {Nandy}, \citenamefont {Nath},\ and\
  \citenamefont {Sahu}}]{Bhattacharya2022}%
  \BibitemOpen
  \bibfield  {author} {\bibinfo {author} {\bibfnamefont {A.}~\bibnamefont
  {Bhattacharya}}, \bibinfo {author} {\bibfnamefont {P.}~\bibnamefont {Nandy}},
  \bibinfo {author} {\bibfnamefont {P.~P.}\ \bibnamefont {Nath}},\ and\
  \bibinfo {author} {\bibfnamefont {H.}~\bibnamefont {Sahu}},\ }\bibfield
  {title} {\bibinfo {title} {Operator growth and krylov construction in
  dissipative open quantum systems},\ }\bibfield  {journal} {\bibinfo
  {journal} {Journal of High Energy Physics}\ }\textbf {\bibinfo {volume}
  {2022}},\ \href {https://doi.org/10.1007/jhep12(2022)081}
  {10.1007/jhep12(2022)081} (\bibinfo {year} {2022})\BibitemShut {NoStop}%
\bibitem [{\citenamefont {Bhattacharya}\ \emph {et~al.}(2023)\citenamefont
  {Bhattacharya}, \citenamefont {Nandy}, \citenamefont {Nath},\ and\
  \citenamefont {Sahu}}]{Bhattacharya2023}%
  \BibitemOpen
  \bibfield  {author} {\bibinfo {author} {\bibfnamefont {A.}~\bibnamefont
  {Bhattacharya}}, \bibinfo {author} {\bibfnamefont {P.}~\bibnamefont {Nandy}},
  \bibinfo {author} {\bibfnamefont {P.~P.}\ \bibnamefont {Nath}},\ and\
  \bibinfo {author} {\bibfnamefont {H.}~\bibnamefont {Sahu}},\ }\bibfield
  {title} {\bibinfo {title} {On krylov complexity in open systems: an approach
  via bi-lanczos algorithm},\ }\bibfield  {journal} {\bibinfo  {journal}
  {Journal of High Energy Physics}\ }\textbf {\bibinfo {volume} {2023}},\ \href
  {https://doi.org/10.1007/jhep12(2023)066} {10.1007/jhep12(2023)066} (\bibinfo
  {year} {2023})\BibitemShut {NoStop}%
\bibitem [{\citenamefont {Menzler}\ and\ \citenamefont
  {Jha}(2024{\natexlab{b}})}]{Menzler2024}%
  \BibitemOpen
  \bibfield  {author} {\bibinfo {author} {\bibfnamefont {H.~G.}\ \bibnamefont
  {Menzler}}\ and\ \bibinfo {author} {\bibfnamefont {R.}~\bibnamefont {Jha}},\
  }\bibfield  {title} {\bibinfo {title} {Krylov delocalization/localization
  across ergodicity breaking},\ }\href
  {https://doi.org/10.1103/PhysRevB.110.125137} {\bibfield  {journal} {\bibinfo
   {journal} {Phys. Rev. B}\ }\textbf {\bibinfo {volume} {110}},\ \bibinfo
  {pages} {125137} (\bibinfo {year} {2024}{\natexlab{b}})}\BibitemShut
  {NoStop}%
\bibitem [{\citenamefont {Malik}\ \emph {et~al.}(2026)\citenamefont {Malik},
  \citenamefont {Sharma}, \citenamefont {Shukla}, \citenamefont {Aravinda},\
  and\ \citenamefont {Mishra}}]{Malik_prb_2026}%
  \BibitemOpen
  \bibfield  {author} {\bibinfo {author} {\bibfnamefont {G.~R.}\ \bibnamefont
  {Malik}}, \bibinfo {author} {\bibfnamefont {J.}~\bibnamefont {Sharma}},
  \bibinfo {author} {\bibfnamefont {R.~K.}\ \bibnamefont {Shukla}}, \bibinfo
  {author} {\bibfnamefont {S.}~\bibnamefont {Aravinda}},\ and\ \bibinfo
  {author} {\bibfnamefont {S.~K.}\ \bibnamefont {Mishra}},\ }\bibfield  {title}
  {\bibinfo {title} {Krylov complexity in the ergodically constrained
  nonintegrable transverse-field ising model},\ }\href
  {https://doi.org/10.1103/45qp-ktk1} {\bibfield  {journal} {\bibinfo
  {journal} {Phys. Rev. B}\ }\textbf {\bibinfo {volume} {113}},\ \bibinfo
  {pages} {184207} (\bibinfo {year} {2026})}\BibitemShut {NoStop}%
\bibitem [{\citenamefont {Balasubramanian}\ \emph {et~al.}(2022)\citenamefont
  {Balasubramanian}, \citenamefont {Caputa}, \citenamefont {Magan},\ and\
  \citenamefont {Wu}}]{Balasubramanian2022}%
  \BibitemOpen
  \bibfield  {author} {\bibinfo {author} {\bibfnamefont {V.}~\bibnamefont
  {Balasubramanian}}, \bibinfo {author} {\bibfnamefont {P.}~\bibnamefont
  {Caputa}}, \bibinfo {author} {\bibfnamefont {J.~M.}\ \bibnamefont {Magan}},\
  and\ \bibinfo {author} {\bibfnamefont {Q.}~\bibnamefont {Wu}},\ }\bibfield
  {title} {\bibinfo {title} {Quantum chaos and the complexity of spread of
  states},\ }\href {https://doi.org/10.1103/PhysRevD.106.046007} {\bibfield
  {journal} {\bibinfo  {journal} {Phys. Rev. D}\ }\textbf {\bibinfo {volume}
  {106}},\ \bibinfo {pages} {046007} (\bibinfo {year} {2022})}\BibitemShut
  {NoStop}%
\bibitem [{\citenamefont {Seetharaman}\ \emph {et~al.}(2025)\citenamefont
  {Seetharaman}, \citenamefont {Singh},\ and\ \citenamefont
  {Nath}}]{Seetharaman_prd_2025}%
  \BibitemOpen
  \bibfield  {author} {\bibinfo {author} {\bibfnamefont {S.}~\bibnamefont
  {Seetharaman}}, \bibinfo {author} {\bibfnamefont {C.}~\bibnamefont {Singh}},\
  and\ \bibinfo {author} {\bibfnamefont {R.}~\bibnamefont {Nath}},\ }\bibfield
  {title} {\bibinfo {title} {Properties of krylov state complexity in qubit
  dynamics},\ }\href {https://doi.org/10.1103/PhysRevD.111.076014} {\bibfield
  {journal} {\bibinfo  {journal} {Phys. Rev. D}\ }\textbf {\bibinfo {volume}
  {111}},\ \bibinfo {pages} {076014} (\bibinfo {year} {2025})}\BibitemShut
  {NoStop}%
\bibitem [{\citenamefont {Bento}\ \emph {et~al.}(2024)\citenamefont {Bento},
  \citenamefont {del Campo},\ and\ \citenamefont {C\'eleri}}]{Bento2024}%
  \BibitemOpen
  \bibfield  {author} {\bibinfo {author} {\bibfnamefont {P.~H.~S.}\
  \bibnamefont {Bento}}, \bibinfo {author} {\bibfnamefont {A.}~\bibnamefont
  {del Campo}},\ and\ \bibinfo {author} {\bibfnamefont {L.~C.}\ \bibnamefont
  {C\'eleri}},\ }\bibfield  {title} {\bibinfo {title} {Krylov complexity and
  dynamical phase transition in the quenched lipkin-meshkov-glick model},\
  }\href {https://doi.org/10.1103/PhysRevB.109.224304} {\bibfield  {journal}
  {\bibinfo  {journal} {Phys. Rev. B}\ }\textbf {\bibinfo {volume} {109}},\
  \bibinfo {pages} {224304} (\bibinfo {year} {2024})}\BibitemShut {NoStop}%
\bibitem [{\citenamefont {Baggioli}\ \emph {et~al.}(2025)\citenamefont
  {Baggioli}, \citenamefont {Huh}, \citenamefont {Jeong}, \citenamefont {Kim},\
  and\ \citenamefont {Pedraza}}]{Baggioli2025}%
  \BibitemOpen
  \bibfield  {author} {\bibinfo {author} {\bibfnamefont {M.}~\bibnamefont
  {Baggioli}}, \bibinfo {author} {\bibfnamefont {K.-B.}\ \bibnamefont {Huh}},
  \bibinfo {author} {\bibfnamefont {H.-S.}\ \bibnamefont {Jeong}}, \bibinfo
  {author} {\bibfnamefont {K.-Y.}\ \bibnamefont {Kim}},\ and\ \bibinfo {author}
  {\bibfnamefont {J.~F.}\ \bibnamefont {Pedraza}},\ }\bibfield  {title}
  {\bibinfo {title} {Krylov complexity as an order parameter for quantum
  chaotic-integrable transitions},\ }\href
  {https://doi.org/10.1103/PhysRevResearch.7.023028} {\bibfield  {journal}
  {\bibinfo  {journal} {Phys. Rev. Res.}\ }\textbf {\bibinfo {volume} {7}},\
  \bibinfo {pages} {023028} (\bibinfo {year} {2025})}\BibitemShut {NoStop}%
\bibitem [{\citenamefont {Balasubramanian}\ \emph {et~al.}(2025)\citenamefont
  {Balasubramanian}, \citenamefont {Magan},\ and\ \citenamefont
  {Wu}}]{Balasubramanian2025}%
  \BibitemOpen
  \bibfield  {author} {\bibinfo {author} {\bibfnamefont {V.}~\bibnamefont
  {Balasubramanian}}, \bibinfo {author} {\bibfnamefont {J.~M.}\ \bibnamefont
  {Magan}},\ and\ \bibinfo {author} {\bibfnamefont {Q.}~\bibnamefont {Wu}},\
  }\bibfield  {title} {\bibinfo {title} {Quantum chaos, integrability, and late
  times in the krylov basis},\ }\href
  {https://doi.org/10.1103/PhysRevE.111.014218} {\bibfield  {journal} {\bibinfo
   {journal} {Phys. Rev. E}\ }\textbf {\bibinfo {volume} {111}},\ \bibinfo
  {pages} {014218} (\bibinfo {year} {2025})}\BibitemShut {NoStop}%
\bibitem [{\citenamefont {Bhattacharjee}\ and\ \citenamefont
  {Nandy}(2025)}]{Bhattacharjee2025}%
  \BibitemOpen
  \bibfield  {author} {\bibinfo {author} {\bibfnamefont {B.}~\bibnamefont
  {Bhattacharjee}}\ and\ \bibinfo {author} {\bibfnamefont {P.}~\bibnamefont
  {Nandy}},\ }\bibfield  {title} {\bibinfo {title} {Krylov fractality and
  complexity in generic random matrix ensembles},\ }\href
  {https://doi.org/10.1103/PhysRevB.111.L060202} {\bibfield  {journal}
  {\bibinfo  {journal} {Phys. Rev. B}\ }\textbf {\bibinfo {volume} {111}},\
  \bibinfo {pages} {L060202} (\bibinfo {year} {2025})}\BibitemShut {NoStop}%
\bibitem [{\citenamefont {Bhattacharya}\ \emph
  {et~al.}(2024{\natexlab{a}})\citenamefont {Bhattacharya}, \citenamefont
  {Nath},\ and\ \citenamefont {Sahu}}]{Bhattacharya2024}%
  \BibitemOpen
  \bibfield  {author} {\bibinfo {author} {\bibfnamefont {A.}~\bibnamefont
  {Bhattacharya}}, \bibinfo {author} {\bibfnamefont {P.~P.}\ \bibnamefont
  {Nath}},\ and\ \bibinfo {author} {\bibfnamefont {H.}~\bibnamefont {Sahu}},\
  }\bibfield  {title} {\bibinfo {title} {Krylov complexity for nonlocal spin
  chains},\ }\href {https://doi.org/10.1103/PhysRevD.109.066010} {\bibfield
  {journal} {\bibinfo  {journal} {Phys. Rev. D}\ }\textbf {\bibinfo {volume}
  {109}},\ \bibinfo {pages} {066010} (\bibinfo {year}
  {2024}{\natexlab{a}})}\BibitemShut {NoStop}%
\bibitem [{\citenamefont {Caputa}\ and\ \citenamefont
  {Liu}(2022)}]{Caputa2022}%
  \BibitemOpen
  \bibfield  {author} {\bibinfo {author} {\bibfnamefont {P.}~\bibnamefont
  {Caputa}}\ and\ \bibinfo {author} {\bibfnamefont {S.}~\bibnamefont {Liu}},\
  }\bibfield  {title} {\bibinfo {title} {Quantum complexity and topological
  phases of matter},\ }\href {https://doi.org/10.1103/PhysRevB.106.195125}
  {\bibfield  {journal} {\bibinfo  {journal} {Phys. Rev. B}\ }\textbf {\bibinfo
  {volume} {106}},\ \bibinfo {pages} {195125} (\bibinfo {year}
  {2022})}\BibitemShut {NoStop}%
\bibitem [{\citenamefont {Bhattacharjee}\ \emph {et~al.}(2022)\citenamefont
  {Bhattacharjee}, \citenamefont {Sur},\ and\ \citenamefont
  {Nandy}}]{Bhattacharjee_prb_2022}%
  \BibitemOpen
  \bibfield  {author} {\bibinfo {author} {\bibfnamefont {B.}~\bibnamefont
  {Bhattacharjee}}, \bibinfo {author} {\bibfnamefont {S.}~\bibnamefont {Sur}},\
  and\ \bibinfo {author} {\bibfnamefont {P.}~\bibnamefont {Nandy}},\ }\bibfield
   {title} {\bibinfo {title} {Probing quantum scars and weak ergodicity
  breaking through quantum complexity},\ }\href
  {https://doi.org/10.1103/PhysRevB.106.205150} {\bibfield  {journal} {\bibinfo
   {journal} {Phys. Rev. B}\ }\textbf {\bibinfo {volume} {106}},\ \bibinfo
  {pages} {205150} (\bibinfo {year} {2022})}\BibitemShut {NoStop}%
\bibitem [{\citenamefont {Takahashi}(2025)}]{Takahashi2025}%
  \BibitemOpen
  \bibfield  {author} {\bibinfo {author} {\bibfnamefont {K.}~\bibnamefont
  {Takahashi}},\ }\bibfield  {title} {\bibinfo {title} {Dynamical quantum phase
  transition, metastable state, and dimensionality reduction: Krylov analysis
  of fully connected spin models},\ }\href {https://doi.org/10.1103/m4jf-7svp}
  {\bibfield  {journal} {\bibinfo  {journal} {Phys. Rev. B}\ }\textbf {\bibinfo
  {volume} {112}},\ \bibinfo {pages} {054312} (\bibinfo {year}
  {2025})}\BibitemShut {NoStop}%
\bibitem [{\citenamefont {Caputa}\ \emph {et~al.}(2023)\citenamefont {Caputa},
  \citenamefont {Gupta}, \citenamefont {Haque}, \citenamefont {Liu},
  \citenamefont {Murugan},\ and\ \citenamefont {Van~Zyl}}]{Caputa2023}%
  \BibitemOpen
  \bibfield  {author} {\bibinfo {author} {\bibfnamefont {P.}~\bibnamefont
  {Caputa}}, \bibinfo {author} {\bibfnamefont {N.}~\bibnamefont {Gupta}},
  \bibinfo {author} {\bibfnamefont {S.~S.}\ \bibnamefont {Haque}}, \bibinfo
  {author} {\bibfnamefont {S.}~\bibnamefont {Liu}}, \bibinfo {author}
  {\bibfnamefont {J.}~\bibnamefont {Murugan}},\ and\ \bibinfo {author}
  {\bibfnamefont {H.~J.~R.}\ \bibnamefont {Van~Zyl}},\ }\bibfield  {title}
  {\bibinfo {title} {Spread complexity and topological transitions in the
  kitaev chain},\ }\bibfield  {journal} {\bibinfo  {journal} {Journal of High
  Energy Physics}\ }\textbf {\bibinfo {volume} {2023}},\ \href
  {https://doi.org/10.1007/jhep01(2023)120} {10.1007/jhep01(2023)120} (\bibinfo
  {year} {2023})\BibitemShut {NoStop}%
\bibitem [{\citenamefont {Pirmoradian}\ \emph {et~al.}(2025)\citenamefont
  {Pirmoradian}, \citenamefont {Sadoogh}, \citenamefont {Teymouri},
  \citenamefont {Abolqasemi-Azad}, \citenamefont {Lahooti},\ and\ \citenamefont
  {Mohammad-Ali}}]{Reza2025}%
  \BibitemOpen
  \bibfield  {author} {\bibinfo {author} {\bibfnamefont {R.}~\bibnamefont
  {Pirmoradian}}, \bibinfo {author} {\bibfnamefont {E.}~\bibnamefont
  {Sadoogh}}, \bibinfo {author} {\bibfnamefont {M.}~\bibnamefont {Teymouri}},
  \bibinfo {author} {\bibfnamefont {N.}~\bibnamefont {Abolqasemi-Azad}},
  \bibinfo {author} {\bibfnamefont {M.~R.}\ \bibnamefont {Lahooti}},\ and\
  \bibinfo {author} {\bibfnamefont {Z.}~\bibnamefont {Mohammad-Ali}},\ }\href
  {https://arxiv.org/abs/2512.21713} {\bibinfo {title} {Investigation of
  quantum chaos in local and non-local ising models}} (\bibinfo {year}
  {2025}),\ \Eprint {https://arxiv.org/abs/2512.21713} {arXiv:2512.21713
  [quant-ph]} \BibitemShut {NoStop}%
\bibitem [{\citenamefont {Balasubramanian}\ \emph {et~al.}(2023)\citenamefont
  {Balasubramanian}, \citenamefont {Magan},\ and\ \citenamefont
  {Wu}}]{Balasubramanian2023}%
  \BibitemOpen
  \bibfield  {author} {\bibinfo {author} {\bibfnamefont {V.}~\bibnamefont
  {Balasubramanian}}, \bibinfo {author} {\bibfnamefont {J.~M.}\ \bibnamefont
  {Magan}},\ and\ \bibinfo {author} {\bibfnamefont {Q.}~\bibnamefont {Wu}},\
  }\bibfield  {title} {\bibinfo {title} {Tridiagonalizing random matrices},\
  }\href {https://doi.org/10.1103/PhysRevD.107.126001} {\bibfield  {journal}
  {\bibinfo  {journal} {Phys. Rev. D}\ }\textbf {\bibinfo {volume} {107}},\
  \bibinfo {pages} {126001} (\bibinfo {year} {2023})}\BibitemShut {NoStop}%
\bibitem [{\citenamefont {Gautam}\ \emph {et~al.}(2024)\citenamefont {Gautam},
  \citenamefont {Pal}, \citenamefont {Pal}, \citenamefont {Gill}, \citenamefont
  {Jaiswal},\ and\ \citenamefont {Sarkar}}]{mamta_prb_2024}%
  \BibitemOpen
  \bibfield  {author} {\bibinfo {author} {\bibfnamefont {M.}~\bibnamefont
  {Gautam}}, \bibinfo {author} {\bibfnamefont {K.}~\bibnamefont {Pal}},
  \bibinfo {author} {\bibfnamefont {K.}~\bibnamefont {Pal}}, \bibinfo {author}
  {\bibfnamefont {A.}~\bibnamefont {Gill}}, \bibinfo {author} {\bibfnamefont
  {N.}~\bibnamefont {Jaiswal}},\ and\ \bibinfo {author} {\bibfnamefont
  {T.}~\bibnamefont {Sarkar}},\ }\bibfield  {title} {\bibinfo {title} {Spread
  complexity evolution in quenched interacting quantum systems},\ }\href
  {https://doi.org/10.1103/PhysRevB.109.014312} {\bibfield  {journal} {\bibinfo
   {journal} {Phys. Rev. B}\ }\textbf {\bibinfo {volume} {109}},\ \bibinfo
  {pages} {014312} (\bibinfo {year} {2024})}\BibitemShut {NoStop}%
\bibitem [{\citenamefont {Zhou}\ \emph {et~al.}(2025)\citenamefont {Zhou},
  \citenamefont {Xia}, \citenamefont {Li},\ and\ \citenamefont
  {Li}}]{zhou_prr_2025}%
  \BibitemOpen
  \bibfield  {author} {\bibinfo {author} {\bibfnamefont {Y.}~\bibnamefont
  {Zhou}}, \bibinfo {author} {\bibfnamefont {W.}~\bibnamefont {Xia}}, \bibinfo
  {author} {\bibfnamefont {L.}~\bibnamefont {Li}},\ and\ \bibinfo {author}
  {\bibfnamefont {W.}~\bibnamefont {Li}},\ }\bibfield  {title} {\bibinfo
  {title} {Diagnosing quantum many-body chaos in non-hermitian quantum spin
  chain via krylov complexity},\ }\href {https://doi.org/10.1103/fw62-j2n9}
  {\bibfield  {journal} {\bibinfo  {journal} {Phys. Rev. Res.}\ }\textbf
  {\bibinfo {volume} {7}},\ \bibinfo {pages} {033281} (\bibinfo {year}
  {2025})}\BibitemShut {NoStop}%
\bibitem [{\citenamefont {Bhattacharya}\ \emph
  {et~al.}(2024{\natexlab{b}})\citenamefont {Bhattacharya}, \citenamefont
  {Das}, \citenamefont {Dey},\ and\ \citenamefont
  {Erdmenger}}]{bhattachrya_prb_2024}%
  \BibitemOpen
  \bibfield  {author} {\bibinfo {author} {\bibfnamefont {A.}~\bibnamefont
  {Bhattacharya}}, \bibinfo {author} {\bibfnamefont {R.~N.}\ \bibnamefont
  {Das}}, \bibinfo {author} {\bibfnamefont {B.}~\bibnamefont {Dey}},\ and\
  \bibinfo {author} {\bibfnamefont {J.}~\bibnamefont {Erdmenger}},\ }\bibfield
  {title} {\bibinfo {title} {Spread complexity and localization in
  $\mathcal{PT}$-symmetric systems},\ }\href
  {https://doi.org/10.1103/PhysRevB.110.064320} {\bibfield  {journal} {\bibinfo
   {journal} {Phys. Rev. B}\ }\textbf {\bibinfo {volume} {110}},\ \bibinfo
  {pages} {064320} (\bibinfo {year} {2024}{\natexlab{b}})}\BibitemShut
  {NoStop}%
\bibitem [{\citenamefont {Medina-Guerra}\ \emph {et~al.}(2025)\citenamefont
  {Medina-Guerra}, \citenamefont {Gornyi},\ and\ \citenamefont
  {Gefen}}]{guerra_prb_2025}%
  \BibitemOpen
  \bibfield  {author} {\bibinfo {author} {\bibfnamefont {E.}~\bibnamefont
  {Medina-Guerra}}, \bibinfo {author} {\bibfnamefont {I.~V.}\ \bibnamefont
  {Gornyi}},\ and\ \bibinfo {author} {\bibfnamefont {Y.}~\bibnamefont
  {Gefen}},\ }\bibfield  {title} {\bibinfo {title} {Correlations and krylov
  spread for a non-hermitian hamiltonian: Ising chain with a complex-valued
  transverse magnetic field},\ }\href
  {https://doi.org/10.1103/PhysRevB.111.174207} {\bibfield  {journal} {\bibinfo
   {journal} {Phys. Rev. B}\ }\textbf {\bibinfo {volume} {111}},\ \bibinfo
  {pages} {174207} (\bibinfo {year} {2025})}\BibitemShut {NoStop}%
\bibitem [{\citenamefont {Nandy}\ \emph
  {et~al.}(2025{\natexlab{b}})\citenamefont {Nandy}, \citenamefont {Pathak},
  \citenamefont {Xian},\ and\ \citenamefont {Erdmenger}}]{Nandy_prb_2025}%
  \BibitemOpen
  \bibfield  {author} {\bibinfo {author} {\bibfnamefont {P.}~\bibnamefont
  {Nandy}}, \bibinfo {author} {\bibfnamefont {T.}~\bibnamefont {Pathak}},
  \bibinfo {author} {\bibfnamefont {Z.-Y.}\ \bibnamefont {Xian}},\ and\
  \bibinfo {author} {\bibfnamefont {J.}~\bibnamefont {Erdmenger}},\ }\bibfield
  {title} {\bibinfo {title} {Krylov space approach to singular value
  decomposition in non-hermitian systems},\ }\href
  {https://doi.org/10.1103/PhysRevB.111.064203} {\bibfield  {journal} {\bibinfo
   {journal} {Phys. Rev. B}\ }\textbf {\bibinfo {volume} {111}},\ \bibinfo
  {pages} {064203} (\bibinfo {year} {2025}{\natexlab{b}})}\BibitemShut
  {NoStop}%
\bibitem [{\citenamefont {Sahu}\ \emph {et~al.}(2026)\citenamefont {Sahu},
  \citenamefont {Bhattacharya},\ and\ \citenamefont {Nath}}]{sahu_prb_2026}%
  \BibitemOpen
  \bibfield  {author} {\bibinfo {author} {\bibfnamefont {H.}~\bibnamefont
  {Sahu}}, \bibinfo {author} {\bibfnamefont {A.}~\bibnamefont {Bhattacharya}},\
  and\ \bibinfo {author} {\bibfnamefont {P.~P.}\ \bibnamefont {Nath}},\
  }\bibfield  {title} {\bibinfo {title} {Quantum complexity and localization in
  random and time-periodic unitary circuits},\ }\href
  {https://doi.org/10.1103/mf2z-mrpd} {\bibfield  {journal} {\bibinfo
  {journal} {Phys. Rev. B}\ }\textbf {\bibinfo {volume} {113}},\ \bibinfo
  {pages} {214312} (\bibinfo {year} {2026})}\BibitemShut {NoStop}%
\bibitem [{\citenamefont {Chaki}\ \emph {et~al.}(2026)\citenamefont {Chaki},
  \citenamefont {Sil}, \citenamefont {Ghosh}, \citenamefont {Sen},\ and\
  \citenamefont {Roy}}]{chaki2026}%
  \BibitemOpen
  \bibfield  {author} {\bibinfo {author} {\bibfnamefont {P.}~\bibnamefont
  {Chaki}}, \bibinfo {author} {\bibfnamefont {A.}~\bibnamefont {Sil}}, \bibinfo
  {author} {\bibfnamefont {P.}~\bibnamefont {Ghosh}}, \bibinfo {author}
  {\bibfnamefont {U.}~\bibnamefont {Sen}},\ and\ \bibinfo {author}
  {\bibfnamefont {S.~S.}\ \bibnamefont {Roy}},\ }\href
  {https://arxiv.org/abs/2605.20350} {\bibinfo {title} {Resource generation and
  dynamical complexities in open random quantum circuits}} (\bibinfo {year}
  {2026}),\ \Eprint {https://arxiv.org/abs/2605.20350} {arXiv:2605.20350
  [quant-ph]} \BibitemShut {NoStop}%
\bibitem [{\citenamefont {PG}\ \emph {et~al.}(2025)\citenamefont {PG},
  \citenamefont {Kannan}, \citenamefont {Modak},\ and\ \citenamefont
  {Aravinda}}]{PG2025}%
  \BibitemOpen
  \bibfield  {author} {\bibinfo {author} {\bibfnamefont {S.}~\bibnamefont
  {PG}}, \bibinfo {author} {\bibfnamefont {J.~B.}\ \bibnamefont {Kannan}},
  \bibinfo {author} {\bibfnamefont {R.}~\bibnamefont {Modak}},\ and\ \bibinfo
  {author} {\bibfnamefont {S.}~\bibnamefont {Aravinda}},\ }\bibfield  {title}
  {\bibinfo {title} {Dependence of krylov complexity saturation on the initial
  operator and state},\ }\href {https://doi.org/10.1103/hsvm-w849} {\bibfield
  {journal} {\bibinfo  {journal} {Phys. Rev. E}\ }\textbf {\bibinfo {volume}
  {112}},\ \bibinfo {pages} {L032203} (\bibinfo {year} {2025})}\BibitemShut
  {NoStop}%
\bibitem [{\citenamefont {Peacock}\ \emph {et~al.}(2026)\citenamefont
  {Peacock}, \citenamefont {Oganesyan},\ and\ \citenamefont
  {Sels}}]{Peacock2026}%
  \BibitemOpen
  \bibfield  {author} {\bibinfo {author} {\bibfnamefont {J.~C.}\ \bibnamefont
  {Peacock}}, \bibinfo {author} {\bibfnamefont {V.}~\bibnamefont {Oganesyan}},\
  and\ \bibinfo {author} {\bibfnamefont {D.}~\bibnamefont {Sels}},\ }\bibfield
  {title} {\bibinfo {title} {Anderson localization: A view from krylov space},\
  }\href {https://doi.org/10.1103/y9y9-gfwc} {\bibfield  {journal} {\bibinfo
  {journal} {Phys. Rev. B}\ }\textbf {\bibinfo {volume} {113}},\ \bibinfo
  {pages} {064204} (\bibinfo {year} {2026})}\BibitemShut {NoStop}%
\bibitem [{\citenamefont {Yeh}\ and\ \citenamefont
  {Mitra}(2026)}]{HsiuChung2026}%
  \BibitemOpen
  \bibfield  {author} {\bibinfo {author} {\bibfnamefont {H.-C.}\ \bibnamefont
  {Yeh}}\ and\ \bibinfo {author} {\bibfnamefont {A.}~\bibnamefont {Mitra}},\
  }\href {https://arxiv.org/abs/2605.24115} {\bibinfo {title} {Anderson
  localization: A floquet operator krylov space perspective}} (\bibinfo {year}
  {2026}),\ \Eprint {https://arxiv.org/abs/2605.24115} {arXiv:2605.24115
  [cond-mat.dis-nn]} \BibitemShut {NoStop}%
\bibitem [{\citenamefont {Aubry}\ and\ \citenamefont
  {Andre}(1980)}]{aubrey_andre}%
  \BibitemOpen
  \bibfield  {author} {\bibinfo {author} {\bibfnamefont {S.}~\bibnamefont
  {Aubry}}\ and\ \bibinfo {author} {\bibfnamefont {G.}~\bibnamefont {Andre}},\
  }\bibfield  {title} {\bibinfo {title} {Analyticity breaking and anderson
  localization},\ }\href@noop {} {\bibfield  {journal} {\bibinfo  {journal}
  {Ann. Israel Phys. Soc. 3, 133}\ } (\bibinfo {year} {1980})}\BibitemShut
  {NoStop}%
\bibitem [{\citenamefont {Harper}(1955)}]{Harper1955}%
  \BibitemOpen
  \bibfield  {author} {\bibinfo {author} {\bibfnamefont {P.~G.}\ \bibnamefont
  {Harper}},\ }\bibfield  {title} {\bibinfo {title} {Single band motion of
  conduction electrons in a uniform magnetic field},\ }\href
  {https://doi.org/10.1088/0370-1298/68/10/304} {\bibfield  {journal} {\bibinfo
   {journal} {Proceedings of the Physical Society. Section A}\ }\textbf
  {\bibinfo {volume} {68}},\ \bibinfo {pages} {874–878} (\bibinfo {year}
  {1955})}\BibitemShut {NoStop}%
\bibitem [{\citenamefont {Deng}\ \emph {et~al.}(2019)\citenamefont {Deng},
  \citenamefont {Ray}, \citenamefont {Sinha}, \citenamefont {Shlyapnikov},\
  and\ \citenamefont {Santos}}]{Deng2019}%
  \BibitemOpen
  \bibfield  {author} {\bibinfo {author} {\bibfnamefont {X.}~\bibnamefont
  {Deng}}, \bibinfo {author} {\bibfnamefont {S.}~\bibnamefont {Ray}}, \bibinfo
  {author} {\bibfnamefont {S.}~\bibnamefont {Sinha}}, \bibinfo {author}
  {\bibfnamefont {G.~V.}\ \bibnamefont {Shlyapnikov}},\ and\ \bibinfo {author}
  {\bibfnamefont {L.}~\bibnamefont {Santos}},\ }\bibfield  {title} {\bibinfo
  {title} {One-dimensional quasicrystals with power-law hopping},\ }\href
  {https://doi.org/10.1103/PhysRevLett.123.025301} {\bibfield  {journal}
  {\bibinfo  {journal} {Phys. Rev. Lett.}\ }\textbf {\bibinfo {volume} {123}},\
  \bibinfo {pages} {025301} (\bibinfo {year} {2019})}\BibitemShut {NoStop}%
\bibitem [{\citenamefont {Hiramoto}\ and\ \citenamefont
  {Kohmoto}(1989)}]{Hiramoto_prb_1989}%
  \BibitemOpen
  \bibfield  {author} {\bibinfo {author} {\bibfnamefont {H.}~\bibnamefont
  {Hiramoto}}\ and\ \bibinfo {author} {\bibfnamefont {M.}~\bibnamefont
  {Kohmoto}},\ }\bibfield  {title} {\bibinfo {title} {Scaling analysis of
  quasiperiodic systems: Generalized harper model},\ }\href
  {https://doi.org/10.1103/PhysRevB.40.8225} {\bibfield  {journal} {\bibinfo
  {journal} {Phys. Rev. B}\ }\textbf {\bibinfo {volume} {40}},\ \bibinfo
  {pages} {8225} (\bibinfo {year} {1989})}\BibitemShut {NoStop}%
\bibitem [{\citenamefont {Ganeshan}\ \emph {et~al.}(2015)\citenamefont
  {Ganeshan}, \citenamefont {Pixley},\ and\ \citenamefont
  {Das~Sarma}}]{Ganeshan2015}%
  \BibitemOpen
  \bibfield  {author} {\bibinfo {author} {\bibfnamefont {S.}~\bibnamefont
  {Ganeshan}}, \bibinfo {author} {\bibfnamefont {J.~H.}\ \bibnamefont
  {Pixley}},\ and\ \bibinfo {author} {\bibfnamefont {S.}~\bibnamefont
  {Das~Sarma}},\ }\bibfield  {title} {\bibinfo {title} {Nearest neighbor tight
  binding models with an exact mobility edge in one dimension},\ }\href
  {https://doi.org/10.1103/PhysRevLett.114.146601} {\bibfield  {journal}
  {\bibinfo  {journal} {Phys. Rev. Lett.}\ }\textbf {\bibinfo {volume} {114}},\
  \bibinfo {pages} {146601} (\bibinfo {year} {2015})}\BibitemShut {NoStop}%
\bibitem [{\citenamefont {Roy}\ and\ \citenamefont {Sharma}(2021)}]{Roy2021}%
  \BibitemOpen
  \bibfield  {author} {\bibinfo {author} {\bibfnamefont {N.}~\bibnamefont
  {Roy}}\ and\ \bibinfo {author} {\bibfnamefont {A.}~\bibnamefont {Sharma}},\
  }\bibfield  {title} {\bibinfo {title} {Fraction of delocalized eigenstates in
  the long-range aubry-andr\'e-harper model},\ }\href
  {https://doi.org/10.1103/PhysRevB.103.075124} {\bibfield  {journal} {\bibinfo
   {journal} {Phys. Rev. B}\ }\textbf {\bibinfo {volume} {103}},\ \bibinfo
  {pages} {075124} (\bibinfo {year} {2021})}\BibitemShut {NoStop}%
\bibitem [{\citenamefont {Biddle}\ \emph {et~al.}(2011)\citenamefont {Biddle},
  \citenamefont {Priour}, \citenamefont {Wang},\ and\ \citenamefont
  {Das~Sarma}}]{Biddle2011}%
  \BibitemOpen
  \bibfield  {author} {\bibinfo {author} {\bibfnamefont {J.}~\bibnamefont
  {Biddle}}, \bibinfo {author} {\bibfnamefont {D.~J.}\ \bibnamefont {Priour}},
  \bibinfo {author} {\bibfnamefont {B.}~\bibnamefont {Wang}},\ and\ \bibinfo
  {author} {\bibfnamefont {S.}~\bibnamefont {Das~Sarma}},\ }\bibfield  {title}
  {\bibinfo {title} {Localization in one-dimensional lattices with
  non-nearest-neighbor hopping: Generalized anderson and aubry-andr\'e
  models},\ }\href {https://doi.org/10.1103/PhysRevB.83.075105} {\bibfield
  {journal} {\bibinfo  {journal} {Phys. Rev. B}\ }\textbf {\bibinfo {volume}
  {83}},\ \bibinfo {pages} {075105} (\bibinfo {year} {2011})}\BibitemShut
  {NoStop}%
\bibitem [{\citenamefont {Biddle}\ \emph {et~al.}(2009)\citenamefont {Biddle},
  \citenamefont {Wang}, \citenamefont {Priour},\ and\ \citenamefont
  {Das~Sarma}}]{Biddle2009}%
  \BibitemOpen
  \bibfield  {author} {\bibinfo {author} {\bibfnamefont {J.}~\bibnamefont
  {Biddle}}, \bibinfo {author} {\bibfnamefont {B.}~\bibnamefont {Wang}},
  \bibinfo {author} {\bibfnamefont {D.~J.}\ \bibnamefont {Priour}},\ and\
  \bibinfo {author} {\bibfnamefont {S.}~\bibnamefont {Das~Sarma}},\ }\bibfield
  {title} {\bibinfo {title} {Localization in one-dimensional incommensurate
  lattices beyond the aubry-andr\'e model},\ }\href
  {https://doi.org/10.1103/PhysRevA.80.021603} {\bibfield  {journal} {\bibinfo
  {journal} {Phys. Rev. A}\ }\textbf {\bibinfo {volume} {80}},\ \bibinfo
  {pages} {021603(R)} (\bibinfo {year} {2009})}\BibitemShut {NoStop}%
\bibitem [{\citenamefont {Biddle}\ and\ \citenamefont
  {Das~Sarma}(2010)}]{Biddle2010}%
  \BibitemOpen
  \bibfield  {author} {\bibinfo {author} {\bibfnamefont {J.}~\bibnamefont
  {Biddle}}\ and\ \bibinfo {author} {\bibfnamefont {S.}~\bibnamefont
  {Das~Sarma}},\ }\bibfield  {title} {\bibinfo {title} {Predicted mobility
  edges in one-dimensional incommensurate optical lattices: An exactly solvable
  model of anderson localization},\ }\href
  {https://doi.org/10.1103/PhysRevLett.104.070601} {\bibfield  {journal}
  {\bibinfo  {journal} {Phys. Rev. Lett.}\ }\textbf {\bibinfo {volume} {104}},\
  \bibinfo {pages} {070601} (\bibinfo {year} {2010})}\BibitemShut {NoStop}%
\bibitem [{\citenamefont {Yang}\ \emph {et~al.}(2017)\citenamefont {Yang},
  \citenamefont {Wang}, \citenamefont {Wang}, \citenamefont {Gao},\ and\
  \citenamefont {Chen}}]{Yang2017}%
  \BibitemOpen
  \bibfield  {author} {\bibinfo {author} {\bibfnamefont {C.}~\bibnamefont
  {Yang}}, \bibinfo {author} {\bibfnamefont {Y.}~\bibnamefont {Wang}}, \bibinfo
  {author} {\bibfnamefont {P.}~\bibnamefont {Wang}}, \bibinfo {author}
  {\bibfnamefont {X.}~\bibnamefont {Gao}},\ and\ \bibinfo {author}
  {\bibfnamefont {S.}~\bibnamefont {Chen}},\ }\bibfield  {title} {\bibinfo
  {title} {Dynamical signature of localization-delocalization transition in a
  one-dimensional incommensurate lattice},\ }\href
  {https://doi.org/10.1103/PhysRevB.95.184201} {\bibfield  {journal} {\bibinfo
  {journal} {Phys. Rev. B}\ }\textbf {\bibinfo {volume} {95}},\ \bibinfo
  {pages} {184201} (\bibinfo {year} {2017})}\BibitemShut {NoStop}%
\bibitem [{\citenamefont {Qing}\ \emph {et~al.}(2026)\citenamefont {Qing},
  \citenamefont {Chen},\ and\ \citenamefont {Zhang}}]{Qing2026}%
  \BibitemOpen
  \bibfield  {author} {\bibinfo {author} {\bibfnamefont {Y.}~\bibnamefont
  {Qing}}, \bibinfo {author} {\bibfnamefont {Y.-Q.}\ \bibnamefont {Chen}},\
  and\ \bibinfo {author} {\bibfnamefont {S.-X.}\ \bibnamefont {Zhang}},\
  }\bibfield  {title} {\bibinfo {title} {Entanglement growth and information
  capacity in a quasiperiodic system with a single-particle mobility edge},\
  }\href {https://doi.org/10.1103/9kdg-m6yg} {\bibfield  {journal} {\bibinfo
  {journal} {Phys. Rev. B}\ }\textbf {\bibinfo {volume} {113}},\ \bibinfo
  {pages} {064308} (\bibinfo {year} {2026})}\BibitemShut {NoStop}%
\bibitem [{\citenamefont {Ye}\ \emph {et~al.}(2024)\citenamefont {Ye},
  \citenamefont {Zhou}, \citenamefont {Khan},\ and\ \citenamefont
  {Xianlong}}]{Ye2024}%
  \BibitemOpen
  \bibfield  {author} {\bibinfo {author} {\bibfnamefont {S.}~\bibnamefont
  {Ye}}, \bibinfo {author} {\bibfnamefont {Z.}~\bibnamefont {Zhou}}, \bibinfo
  {author} {\bibfnamefont {N.~A.}\ \bibnamefont {Khan}},\ and\ \bibinfo
  {author} {\bibfnamefont {G.}~\bibnamefont {Xianlong}},\ }\bibfield  {title}
  {\bibinfo {title} {Energy-dependent dynamical quantum phase transitions in
  quasicrystals},\ }\href {https://doi.org/10.1103/PhysRevA.109.043319}
  {\bibfield  {journal} {\bibinfo  {journal} {Phys. Rev. A}\ }\textbf {\bibinfo
  {volume} {109}},\ \bibinfo {pages} {043319} (\bibinfo {year}
  {2024})}\BibitemShut {NoStop}%
\bibitem [{\citenamefont {Sachdev}(2011)}]{Sachdev2011}%
  \BibitemOpen
  \bibfield  {author} {\bibinfo {author} {\bibfnamefont {S.}~\bibnamefont
  {Sachdev}},\ }\href {https://doi.org/10.1017/cbo9780511973765} {\emph
  {\bibinfo {title} {Quantum Phase Transitions}}}\ (\bibinfo  {publisher}
  {Cambridge University Press},\ \bibinfo {year} {2011})\BibitemShut {NoStop}%
\bibitem [{\citenamefont {Lye}\ \emph {et~al.}(2005)\citenamefont {Lye},
  \citenamefont {Fallani}, \citenamefont {Modugno}, \citenamefont {Wiersma},
  \citenamefont {Fort},\ and\ \citenamefont {Inguscio}}]{Lye2005}%
  \BibitemOpen
  \bibfield  {author} {\bibinfo {author} {\bibfnamefont {J.~E.}\ \bibnamefont
  {Lye}}, \bibinfo {author} {\bibfnamefont {L.}~\bibnamefont {Fallani}},
  \bibinfo {author} {\bibfnamefont {M.}~\bibnamefont {Modugno}}, \bibinfo
  {author} {\bibfnamefont {D.~S.}\ \bibnamefont {Wiersma}}, \bibinfo {author}
  {\bibfnamefont {C.}~\bibnamefont {Fort}},\ and\ \bibinfo {author}
  {\bibfnamefont {M.}~\bibnamefont {Inguscio}},\ }\bibfield  {title} {\bibinfo
  {title} {Bose-einstein condensate in a random potential},\ }\href
  {https://doi.org/10.1103/PhysRevLett.95.070401} {\bibfield  {journal}
  {\bibinfo  {journal} {Phys. Rev. Lett.}\ }\textbf {\bibinfo {volume} {95}},\
  \bibinfo {pages} {070401} (\bibinfo {year} {2005})}\BibitemShut {NoStop}%
\bibitem [{\citenamefont {Roati}\ \emph {et~al.}(2008)\citenamefont {Roati},
  \citenamefont {D’Errico}, \citenamefont {Fallani}, \citenamefont {Fattori},
  \citenamefont {Fort}, \citenamefont {Zaccanti}, \citenamefont {Modugno},
  \citenamefont {Modugno},\ and\ \citenamefont {Inguscio}}]{Roati2008}%
  \BibitemOpen
  \bibfield  {author} {\bibinfo {author} {\bibfnamefont {G.}~\bibnamefont
  {Roati}}, \bibinfo {author} {\bibfnamefont {C.}~\bibnamefont {D’Errico}},
  \bibinfo {author} {\bibfnamefont {L.}~\bibnamefont {Fallani}}, \bibinfo
  {author} {\bibfnamefont {M.}~\bibnamefont {Fattori}}, \bibinfo {author}
  {\bibfnamefont {C.}~\bibnamefont {Fort}}, \bibinfo {author} {\bibfnamefont
  {M.}~\bibnamefont {Zaccanti}}, \bibinfo {author} {\bibfnamefont
  {G.}~\bibnamefont {Modugno}}, \bibinfo {author} {\bibfnamefont
  {M.}~\bibnamefont {Modugno}},\ and\ \bibinfo {author} {\bibfnamefont
  {M.}~\bibnamefont {Inguscio}},\ }\bibfield  {title} {\bibinfo {title}
  {Anderson localization of a non-interacting bose–einstein condensate},\
  }\href {https://doi.org/10.1038/nature07071} {\bibfield  {journal} {\bibinfo
  {journal} {Nature}\ }\textbf {\bibinfo {volume} {453}},\ \bibinfo {pages}
  {895–898} (\bibinfo {year} {2008})}\BibitemShut {NoStop}%
\bibitem [{\citenamefont {Lahini}\ \emph {et~al.}(2009)\citenamefont {Lahini},
  \citenamefont {Pugatch}, \citenamefont {Pozzi}, \citenamefont {Sorel},
  \citenamefont {Morandotti}, \citenamefont {Davidson},\ and\ \citenamefont
  {Silberberg}}]{Lahini2009}%
  \BibitemOpen
  \bibfield  {author} {\bibinfo {author} {\bibfnamefont {Y.}~\bibnamefont
  {Lahini}}, \bibinfo {author} {\bibfnamefont {R.}~\bibnamefont {Pugatch}},
  \bibinfo {author} {\bibfnamefont {F.}~\bibnamefont {Pozzi}}, \bibinfo
  {author} {\bibfnamefont {M.}~\bibnamefont {Sorel}}, \bibinfo {author}
  {\bibfnamefont {R.}~\bibnamefont {Morandotti}}, \bibinfo {author}
  {\bibfnamefont {N.}~\bibnamefont {Davidson}},\ and\ \bibinfo {author}
  {\bibfnamefont {Y.}~\bibnamefont {Silberberg}},\ }\bibfield  {title}
  {\bibinfo {title} {Observation of a localization transition in quasiperiodic
  photonic lattices},\ }\href {https://doi.org/10.1103/PhysRevLett.103.013901}
  {\bibfield  {journal} {\bibinfo  {journal} {Phys. Rev. Lett.}\ }\textbf
  {\bibinfo {volume} {103}},\ \bibinfo {pages} {013901} (\bibinfo {year}
  {2009})}\BibitemShut {NoStop}%
\bibitem [{\citenamefont {Schreiber}\ \emph {et~al.}(2015)\citenamefont
  {Schreiber}, \citenamefont {Hodgman}, \citenamefont {Bordia}, \citenamefont
  {L\"{u}schen}, \citenamefont {Fischer}, \citenamefont {Vosk}, \citenamefont
  {Altman}, \citenamefont {Schneider},\ and\ \citenamefont
  {Bloch}}]{Schreiber2015}%
  \BibitemOpen
  \bibfield  {author} {\bibinfo {author} {\bibfnamefont {M.}~\bibnamefont
  {Schreiber}}, \bibinfo {author} {\bibfnamefont {S.~S.}\ \bibnamefont
  {Hodgman}}, \bibinfo {author} {\bibfnamefont {P.}~\bibnamefont {Bordia}},
  \bibinfo {author} {\bibfnamefont {H.~P.}\ \bibnamefont {L\"{u}schen}},
  \bibinfo {author} {\bibfnamefont {M.~H.}\ \bibnamefont {Fischer}}, \bibinfo
  {author} {\bibfnamefont {R.}~\bibnamefont {Vosk}}, \bibinfo {author}
  {\bibfnamefont {E.}~\bibnamefont {Altman}}, \bibinfo {author} {\bibfnamefont
  {U.}~\bibnamefont {Schneider}},\ and\ \bibinfo {author} {\bibfnamefont
  {I.}~\bibnamefont {Bloch}},\ }\bibfield  {title} {\bibinfo {title}
  {Observation of many-body localization of interacting fermions in a
  quasirandom optical lattice},\ }\href
  {https://doi.org/10.1126/science.aaa7432} {\bibfield  {journal} {\bibinfo
  {journal} {Science}\ }\textbf {\bibinfo {volume} {349}},\ \bibinfo {pages}
  {842–845} (\bibinfo {year} {2015})}\BibitemShut {NoStop}%
\bibitem [{\citenamefont {Kim}\ \emph {et~al.}(2010)\citenamefont {Kim},
  \citenamefont {Chang}, \citenamefont {Korenblit}, \citenamefont {Islam},
  \citenamefont {Edwards}, \citenamefont {Freericks}, \citenamefont {Lin},
  \citenamefont {Duan},\ and\ \citenamefont {Monroe}}]{Kim2010}%
  \BibitemOpen
  \bibfield  {author} {\bibinfo {author} {\bibfnamefont {K.}~\bibnamefont
  {Kim}}, \bibinfo {author} {\bibfnamefont {M.-S.}\ \bibnamefont {Chang}},
  \bibinfo {author} {\bibfnamefont {S.}~\bibnamefont {Korenblit}}, \bibinfo
  {author} {\bibfnamefont {R.}~\bibnamefont {Islam}}, \bibinfo {author}
  {\bibfnamefont {E.~E.}\ \bibnamefont {Edwards}}, \bibinfo {author}
  {\bibfnamefont {J.~K.}\ \bibnamefont {Freericks}}, \bibinfo {author}
  {\bibfnamefont {G.-D.}\ \bibnamefont {Lin}}, \bibinfo {author} {\bibfnamefont
  {L.-M.}\ \bibnamefont {Duan}},\ and\ \bibinfo {author} {\bibfnamefont
  {C.}~\bibnamefont {Monroe}},\ }\bibfield  {title} {\bibinfo {title} {Quantum
  simulation of frustrated ising spins with trapped ions},\ }\href
  {https://doi.org/10.1038/nature09071} {\bibfield  {journal} {\bibinfo
  {journal} {Nature}\ }\textbf {\bibinfo {volume} {465}},\ \bibinfo {pages}
  {590–593} (\bibinfo {year} {2010})}\BibitemShut {NoStop}%
\bibitem [{\citenamefont {Richerme}\ \emph {et~al.}(2014)\citenamefont
  {Richerme}, \citenamefont {Gong}, \citenamefont {Lee}, \citenamefont {Senko},
  \citenamefont {Smith}, \citenamefont {Foss-Feig}, \citenamefont {Michalakis},
  \citenamefont {Gorshkov},\ and\ \citenamefont {Monroe}}]{Richerme2014}%
  \BibitemOpen
  \bibfield  {author} {\bibinfo {author} {\bibfnamefont {P.}~\bibnamefont
  {Richerme}}, \bibinfo {author} {\bibfnamefont {Z.-X.}\ \bibnamefont {Gong}},
  \bibinfo {author} {\bibfnamefont {A.}~\bibnamefont {Lee}}, \bibinfo {author}
  {\bibfnamefont {C.}~\bibnamefont {Senko}}, \bibinfo {author} {\bibfnamefont
  {J.}~\bibnamefont {Smith}}, \bibinfo {author} {\bibfnamefont
  {M.}~\bibnamefont {Foss-Feig}}, \bibinfo {author} {\bibfnamefont
  {S.}~\bibnamefont {Michalakis}}, \bibinfo {author} {\bibfnamefont {A.~V.}\
  \bibnamefont {Gorshkov}},\ and\ \bibinfo {author} {\bibfnamefont
  {C.}~\bibnamefont {Monroe}},\ }\bibfield  {title} {\bibinfo {title}
  {Non-local propagation of correlations in quantum systems with long-range
  interactions},\ }\href {https://doi.org/10.1038/nature13450} {\bibfield
  {journal} {\bibinfo  {journal} {Nature}\ }\textbf {\bibinfo {volume} {511}},\
  \bibinfo {pages} {198–201} (\bibinfo {year} {2014})}\BibitemShut {NoStop}%
\bibitem [{\citenamefont {Labuhn}\ \emph {et~al.}(2016)\citenamefont {Labuhn},
  \citenamefont {Barredo}, \citenamefont {Ravets}, \citenamefont
  {de~Léséleuc}, \citenamefont {Macrì}, \citenamefont {Lahaye},\ and\
  \citenamefont {Browaeys}}]{Labuhn2016}%
  \BibitemOpen
  \bibfield  {author} {\bibinfo {author} {\bibfnamefont {H.}~\bibnamefont
  {Labuhn}}, \bibinfo {author} {\bibfnamefont {D.}~\bibnamefont {Barredo}},
  \bibinfo {author} {\bibfnamefont {S.}~\bibnamefont {Ravets}}, \bibinfo
  {author} {\bibfnamefont {S.}~\bibnamefont {de~Léséleuc}}, \bibinfo {author}
  {\bibfnamefont {T.}~\bibnamefont {Macrì}}, \bibinfo {author} {\bibfnamefont
  {T.}~\bibnamefont {Lahaye}},\ and\ \bibinfo {author} {\bibfnamefont
  {A.}~\bibnamefont {Browaeys}},\ }\bibfield  {title} {\bibinfo {title}
  {Tunable two-dimensional arrays of single rydberg atoms for realizing quantum
  ising models},\ }\href {https://doi.org/10.1038/nature18274} {\bibfield
  {journal} {\bibinfo  {journal} {Nature}\ }\textbf {\bibinfo {volume} {534}},\
  \bibinfo {pages} {667–670} (\bibinfo {year} {2016})}\BibitemShut {NoStop}%
\bibitem [{\citenamefont {Sahoo}\ \emph {et~al.}(2024)\citenamefont {Sahoo},
  \citenamefont {Mishra},\ and\ \citenamefont {Rakshit}}]{sahoo_pra_2024}%
  \BibitemOpen
  \bibfield  {author} {\bibinfo {author} {\bibfnamefont {A.}~\bibnamefont
  {Sahoo}}, \bibinfo {author} {\bibfnamefont {U.}~\bibnamefont {Mishra}},\ and\
  \bibinfo {author} {\bibfnamefont {D.}~\bibnamefont {Rakshit}},\ }\bibfield
  {title} {\bibinfo {title} {Localization-driven quantum sensing},\ }\href
  {https://doi.org/10.1103/PhysRevA.109.L030601} {\bibfield  {journal}
  {\bibinfo  {journal} {Phys. Rev. A}\ }\textbf {\bibinfo {volume} {109}},\
  \bibinfo {pages} {L030601} (\bibinfo {year} {2024})}\BibitemShut {NoStop}%
\bibitem [{\citenamefont {Sahoo}\ \emph {et~al.}(2025)\citenamefont {Sahoo},
  \citenamefont {Saha},\ and\ \citenamefont {Rakshit}}]{sahoo_prb_2025}%
  \BibitemOpen
  \bibfield  {author} {\bibinfo {author} {\bibfnamefont {A.}~\bibnamefont
  {Sahoo}}, \bibinfo {author} {\bibfnamefont {A.}~\bibnamefont {Saha}},\ and\
  \bibinfo {author} {\bibfnamefont {D.}~\bibnamefont {Rakshit}},\ }\bibfield
  {title} {\bibinfo {title} {Stark localization near aubry-andr\'e
  criticality},\ }\href {https://doi.org/10.1103/PhysRevB.111.024205}
  {\bibfield  {journal} {\bibinfo  {journal} {Phys. Rev. B}\ }\textbf {\bibinfo
  {volume} {111}},\ \bibinfo {pages} {024205} (\bibinfo {year}
  {2025})}\BibitemShut {NoStop}%
\bibitem [{\citenamefont {Ganguli}\ and\ \citenamefont
  {Jana}(2024)}]{Ganguli2024}%
  \BibitemOpen
  \bibfield  {author} {\bibinfo {author} {\bibfnamefont {M.}~\bibnamefont
  {Ganguli}}\ and\ \bibinfo {author} {\bibfnamefont {A.}~\bibnamefont {Jana}},\
  }\href {https://doi.org/10.48550/ARXIV.2409.02186} {\bibinfo {title} {State
  dependent spread complexity dynamics in many-body localization transition}}
  (\bibinfo {year} {2024})\BibitemShut {NoStop}%
\bibitem [{\citenamefont {Scialchi}\ \emph {et~al.}(2024)\citenamefont
  {Scialchi}, \citenamefont {Roncaglia},\ and\ \citenamefont
  {Wisniacki}}]{Scialchi2024}%
  \BibitemOpen
  \bibfield  {author} {\bibinfo {author} {\bibfnamefont {G.~F.}\ \bibnamefont
  {Scialchi}}, \bibinfo {author} {\bibfnamefont {A.~J.}\ \bibnamefont
  {Roncaglia}},\ and\ \bibinfo {author} {\bibfnamefont {D.~A.}\ \bibnamefont
  {Wisniacki}},\ }\bibfield  {title} {\bibinfo {title} {Integrability-to-chaos
  transition through the krylov approach for state evolution},\ }\href
  {https://doi.org/10.1103/PhysRevE.109.054209} {\bibfield  {journal} {\bibinfo
   {journal} {Phys. Rev. E}\ }\textbf {\bibinfo {volume} {109}},\ \bibinfo
  {pages} {054209} (\bibinfo {year} {2024})}\BibitemShut {NoStop}%
\bibitem [{\citenamefont {Camargo}\ \emph {et~al.}(2024)\citenamefont
  {Camargo}, \citenamefont {Huh}, \citenamefont {Jahnke}, \citenamefont
  {Jeong}, \citenamefont {Kim},\ and\ \citenamefont {Nishida}}]{Camargo2024}%
  \BibitemOpen
  \bibfield  {author} {\bibinfo {author} {\bibfnamefont {H.~A.}\ \bibnamefont
  {Camargo}}, \bibinfo {author} {\bibfnamefont {K.-B.}\ \bibnamefont {Huh}},
  \bibinfo {author} {\bibfnamefont {V.}~\bibnamefont {Jahnke}}, \bibinfo
  {author} {\bibfnamefont {H.-S.}\ \bibnamefont {Jeong}}, \bibinfo {author}
  {\bibfnamefont {K.-Y.}\ \bibnamefont {Kim}},\ and\ \bibinfo {author}
  {\bibfnamefont {M.}~\bibnamefont {Nishida}},\ }\bibfield  {title} {\bibinfo
  {title} {Spread and spectral complexity in quantum spin chains: from
  integrability to chaos},\ }\bibfield  {journal} {\bibinfo  {journal} {Journal
  of High Energy Physics}\ }\textbf {\bibinfo {volume} {2024}},\ \href
  {https://doi.org/10.1007/jhep08(2024)241} {10.1007/jhep08(2024)241} (\bibinfo
  {year} {2024})\BibitemShut {NoStop}%
\bibitem [{\citenamefont {Grabarits}\ and\ \citenamefont {del
  Campo}(2025)}]{Grabarits2025}%
  \BibitemOpen
  \bibfield  {author} {\bibinfo {author} {\bibfnamefont {A.}~\bibnamefont
  {Grabarits}}\ and\ \bibinfo {author} {\bibfnamefont {A.}~\bibnamefont {del
  Campo}},\ }\href {https://doi.org/10.48550/ARXIV.2510.13947} {\bibinfo
  {title} {Universal growth of krylov complexity across a quantum phase
  transition}} (\bibinfo {year} {2025})\BibitemShut {NoStop}%
\bibitem [{\citenamefont {Teh}\ and\ \citenamefont {Orito}(2025)}]{Teh2025}%
  \BibitemOpen
  \bibfield  {author} {\bibinfo {author} {\bibfnamefont {H.-H.}\ \bibnamefont
  {Teh}}\ and\ \bibinfo {author} {\bibfnamefont {T.}~\bibnamefont {Orito}},\
  }\href {https://doi.org/10.48550/ARXIV.2510.22542} {\bibinfo {title} {Krylov
  complexity and mixed-state phase transition}} (\bibinfo {year}
  {2025})\BibitemShut {NoStop}%
\bibitem [{\citenamefont {Heyl}(2018)}]{Heyl2018}%
  \BibitemOpen
  \bibfield  {author} {\bibinfo {author} {\bibfnamefont {M.}~\bibnamefont
  {Heyl}},\ }\bibfield  {title} {\bibinfo {title} {Dynamical quantum phase
  transitions: a review},\ }\href {https://doi.org/10.1088/1361-6633/aaaf9a}
  {\bibfield  {journal} {\bibinfo  {journal} {Reports on Progress in Physics}\
  }\textbf {\bibinfo {volume} {81}},\ \bibinfo {pages} {054001} (\bibinfo
  {year} {2018})}\BibitemShut {NoStop}%
\bibitem [{\citenamefont {Shimasaki}\ \emph {et~al.}(2022)\citenamefont
  {Shimasaki}, \citenamefont {Kondakci}, \citenamefont {Prichard},
  \citenamefont {Pagett}, \citenamefont {Bai}, \citenamefont {Dotti},
  \citenamefont {Cao}, \citenamefont {Lu}, \citenamefont {Grover},\ and\
  \citenamefont {Weld}}]{Shimasaki2022}%
  \BibitemOpen
  \bibfield  {author} {\bibinfo {author} {\bibfnamefont {T.}~\bibnamefont
  {Shimasaki}}, \bibinfo {author} {\bibfnamefont {H.~E.}\ \bibnamefont
  {Kondakci}}, \bibinfo {author} {\bibfnamefont {M.}~\bibnamefont {Prichard}},
  \bibinfo {author} {\bibfnamefont {J.}~\bibnamefont {Pagett}}, \bibinfo
  {author} {\bibfnamefont {Y.}~\bibnamefont {Bai}}, \bibinfo {author}
  {\bibfnamefont {P.}~\bibnamefont {Dotti}}, \bibinfo {author} {\bibfnamefont
  {A.}~\bibnamefont {Cao}}, \bibinfo {author} {\bibfnamefont {T.-C.}\
  \bibnamefont {Lu}}, \bibinfo {author} {\bibfnamefont {T.}~\bibnamefont
  {Grover}},\ and\ \bibinfo {author} {\bibfnamefont {D.~M.}\ \bibnamefont
  {Weld}},\ }\bibfield  {title} {\bibinfo {title} {Experimental realization of
  the kicked aubry-andre-harper hamiltonian},\ }in\ \href
  {https://doi.org/10.1364/cleo_qels.2022.fm4d.3} {\emph {\bibinfo {booktitle}
  {Conference on Lasers and Electro-Optics}}},\ \bibinfo {series and number}
  {CLEO QELS}\ (\bibinfo  {publisher} {Optica Publishing Group},\ \bibinfo
  {year} {2022})\ p.\ \bibinfo {pages} {FM4D.3}\BibitemShut {NoStop}%
\bibitem [{\citenamefont {Domínguez-Castro}\ and\ \citenamefont
  {Paredes}(2019)}]{DomnguezCastro2019}%
  \BibitemOpen
  \bibfield  {author} {\bibinfo {author} {\bibfnamefont {G.~A.}\ \bibnamefont
  {Domínguez-Castro}}\ and\ \bibinfo {author} {\bibfnamefont {R.}~\bibnamefont
  {Paredes}},\ }\bibfield  {title} {\bibinfo {title} {The aubry–andré model
  as a hobbyhorse for understanding the localization phenomenon},\ }\href
  {https://doi.org/10.1088/1361-6404/ab1670} {\bibfield  {journal} {\bibinfo
  {journal} {European Journal of Physics}\ }\textbf {\bibinfo {volume} {40}},\
  \bibinfo {pages} {045403} (\bibinfo {year} {2019})}\BibitemShut {NoStop}%
\bibitem [{Note1()}]{Note1}%
  \BibitemOpen
  \bibinfo {note} {We also notice that the double derivative of \(\protect
  \mathcal {C}\) with respect to \(\lambda _f\) exhibits the sharp kinks,
  thereby capable of capturing these mobility edges in the gAAH
  model.}\BibitemShut {Stop}%
\bibitem [{\citenamefont {Bhattacharya}\ \emph
  {et~al.}(2024{\natexlab{c}})\citenamefont {Bhattacharya}, \citenamefont
  {Nath},\ and\ \citenamefont {Sahu}}]{Bhattacharya_prd_2024}%
  \BibitemOpen
  \bibfield  {author} {\bibinfo {author} {\bibfnamefont {A.}~\bibnamefont
  {Bhattacharya}}, \bibinfo {author} {\bibfnamefont {P.~P.}\ \bibnamefont
  {Nath}},\ and\ \bibinfo {author} {\bibfnamefont {H.}~\bibnamefont {Sahu}},\
  }\bibfield  {title} {\bibinfo {title} {Krylov complexity for nonlocal spin
  chains},\ }\href {https://doi.org/10.1103/PhysRevD.109.066010} {\bibfield
  {journal} {\bibinfo  {journal} {Phys. Rev. D}\ }\textbf {\bibinfo {volume}
  {109}},\ \bibinfo {pages} {066010} (\bibinfo {year}
  {2024}{\natexlab{c}})}\BibitemShut {NoStop}%
\end{thebibliography}%
\end{document}